\documentclass[final,hidelinks]{siamart251216}
\usepackage{graphicx}
\usepackage{epstopdf}
\usepackage{epsfig}
\usepackage{overpic}
\usepackage{array}
\usepackage{color}
\usepackage{multirow}
\usepackage{float}
\usepackage[caption=false]{subfig}
\usepackage{amsmath,amssymb}
\usepackage[mathscr]{euscript}
\usepackage{latexsym,bm}
\usepackage{booktabs}
\usepackage{bbm}
 \usepackage[all]{xy}
 \usepackage{color}
\usepackage{diagbox}
\usepackage{algorithm}
\usepackage{algorithmicx}
\usepackage{algpseudocode}
\usepackage{hyperref}
\usepackage{makecell}
\usepackage{mathtools}
\usepackage{multirow}
\usepackage{dsfont}
\usepackage[section]{placeins}

\usepackage{physics}

\renewtheorem{theorem}{Theorem}
\renewtheorem{lemma}{Lemma}
\renewtheorem{proposition}{Proposition}
\newtheorem{remark}{Remark}

\newcommand\comment[1]{}
\newcommand{\Var}{\text{Var}}

\def\me{\mathbbm{e}}
\def\D{\textup{d}}
\def\mone{\mathbbm{1}}

\makeatletter
\@addtoreset{equation}{section}
\makeatother
\renewcommand{\theequation}{\arabic{section}.\arabic{equation}}

\graphicspath{ {} }

\newsavebox{\tablebox}

\headers{Particle annihilation in long-time Wigner dynamics}{J. Huang, S. Shao, and Y. Zhang}

\begin{document}
\bibliographystyle{siamplain}

\title{Efficient scheduling of particle annihilation for long-time Wigner simulations}
\author{Jingyang Huang\footnotemark[1], Sihong Shao\footnotemark[1], \and Yuxuan Zhang\footnotemark[1] %\footnotemark[2]
}
\renewcommand{\thefootnote}{\fnsymbol{footnote}}
\footnotetext[1]{CAPT, LMAM and School of Mathematical Sciences, Peking University, Beijing 100871, China. Email addresses:{\tt jyhuang03@stu.pku.edu.cn} (J. Huang) {\tt sihong@math.pku.edu.cn} (S. Shao) {\tt zhangyx@stu.pku.edu.cn} (Y. Zhang).}
\maketitle

% .......... the text...................

\begin{abstract}
Particle annihilation (PA) plays an indispensable role in alleviating the numerical sign problem in long-time simulations with the Wigner branching random walk (WBRW) method. However, the effect of PA frequency on long-time accuracy remains poorly understood, leaving its choice largely empirical. To provide systematic guidance for PA scheduling, we establish an overall error estimate for PA through a uniform mesh, separating WBRW variance, resampling variance and squared bias. The analysis shows that, beyond merely controlling particle growth, PA replaces the exponential-in-time variance bound with a linear-in-time bound, thereby alleviating the numerical sign problem. It also reveals bias-variance trade-off induced by PA frequency: frequent PA limits variance but can increase accumulated bias. More importantly, building on the newfound error structure, we propose a variance-constrained cost optimization strategy for adaptive PA scheduling across several PA methods. An intrinsic-time reparameterization provides a common clock for variance control, PA scheduling, and branching dynamics under time-dependent potentials. Numerical experiments validate the theoretical variance behavior and bias-variance trade-off, and results of our schedule populate the low-cost regime of the empirical Pareto frontier objecting to balance accuracy and efficiency.
\end{abstract}

\begin{keywords}
Sign problem;
Particle annihilation;
Wigner equation;
Long-time simulation;
Variance-constrained cost optimization;
Nonlocal operator.
\end{keywords}
\begin{MSCcodes}
65C35, %Stochastic particle methods
65Y20, %Complexity and performance of numerical algorithms
65M15, %Error bounds for initial value and initial-boundary value problems involving PDEs
62G09, %Nonparametric statistical resampling methods
81S30. %Phase-space methods including Wigner distributions, etc. applied to problems in quantum mechanics
\end{MSCcodes}

\section{Introduction}
\label{sec:intro}
The Wigner function and its governing equation provide an equivalent formulation of quantum mechanics in phase space~\cite{Wigner1932wigner}, which is inherently high-dimensional, leading to computational challenges due to the curse of dimensionality (CoD). To alleviate CoD, a class of particle-based stochastic methods has been developed~\cite{Kosina2003,Kosina2006,Querlioz2010MC, Sellier2015MC}. Among them, the Wigner branching random walk (WBRW) method has been proposed and applied to numerical simulations in phase spaces of up to 12 dimensions~\cite{Shao2019, Xiong2019, Xiong2024SPADE}. In all these methods, the infamous numerical sign problem poses a fundamental obstacle, manifesting as exponential growth in both the particle number and stochastic variance induced by increments of signed weights~\cite{Xiong2024SPADE, Schmidt1993signproblem, Nedjalkov2003sign, Shao2020}. Particle annihilation (PA) is thus widely employed to mitigate the problem and plays an indispensable role in long-time simulations~\cite{Kosina2003, Kosina2006}. However, numerical experiments show that the numerical error and computational cost are sensitive to PA frequency~\cite{Shao2019,Xiong2019}, raising the question: 
\begin{center}
    \vspace{0.2cm}
    \emph{How should PA be scheduled to balance accuracy and efficiency?}
    \vspace{0.2cm}
\end{center}

\begin{figure}[H]
  \centering
  \subfloat[Particle number]{%
    \includegraphics[width=0.32\textwidth]{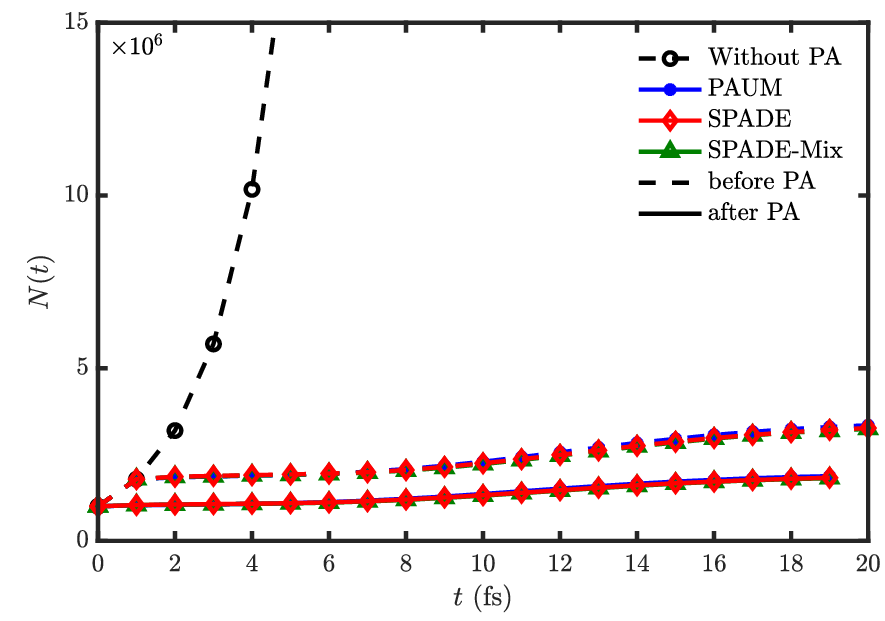}}\hspace{0.5cm}
  \subfloat[Relative error]{%
    \includegraphics[width=0.32\textwidth]{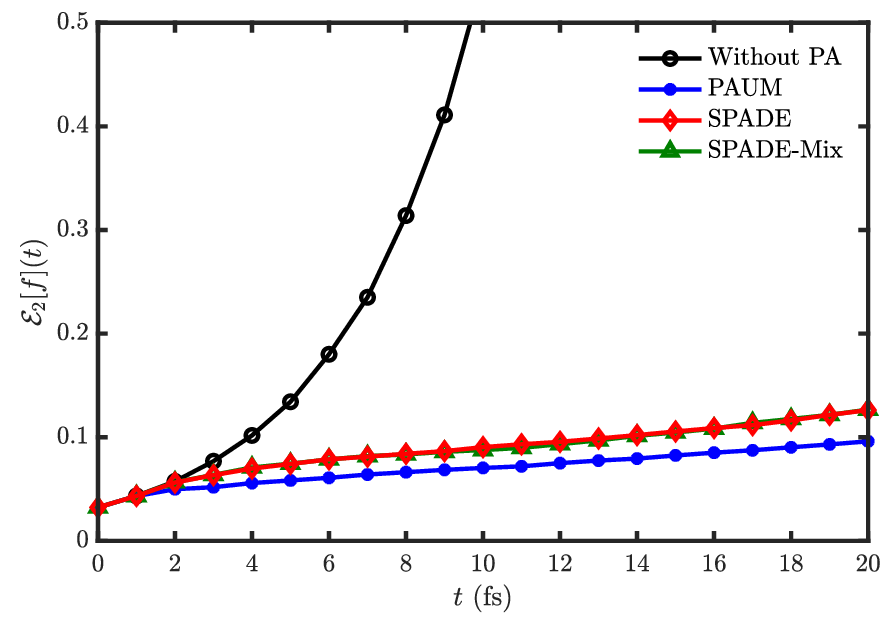}}
  \caption{Particle number $N(t)$ and relative error $\mathcal{E}_{2}[f](t)$ over time obtained by Without PA, PAUM, SPADE, SPADE-Mix at $t=1,\cdots,20$ fs at fixed PA interval $T_A=1$ fs, under the same setting of Section~\ref{sec:as1}.} 
	\label{fig:pa-effect}
\end{figure}

The basic idea of PA is to cluster particles according to domain partition, and to cancel contributions of particles with opposite weights. Particle annihilation through a uniform mesh (PAUM)~\cite{Sellier2014} is a common approach, borrowing the idea of histogram density estimation. Also, several PA methods have been developed for the scalability in high-dimensional problems, such as the sequential-clustering particle annihilation via star discrepancy estimation (SPADE) \cite{Xiong2024SPADE, Shao2020SPADE, Li2016}, and SPADE-Mix which replaces star discrepancy with mixture discrepancy as the uniformity measure \cite{Lei2025discrepancy}. To illustrate the necessity of PA for long-time Wigner simulations, we compare the numerical results without PA and with each of the three PA methods in Figure~\ref{fig:pa-effect}. It can be found that without PA, both particle number and error grow rapidly, while those with PAUM, SPADE and SPADE-Mix remain relatively stable. 

Typically, PA is performed at fixed frequency, and the choice of PA frequency is an important practical issue. In addition to influencing the particle number and numerical error, PA itself can be computationally expensive, especially when adaptive partitioning is employed or in high-dimensional problems~\cite{Lei2025discrepancy}. Despite its practical importance, the influence of PA frequency on the overall numerical error has not yet been quantitatively characterized. The difficulty is to track how errors introduced by repeated PA propagate over time, and how they interact with the stochastic variance from WBRW between successive PA steps. Addressing the scheduling question therefore requires answering the following theoretical question:
\begin{center}
    \vspace{0.2cm}
    \emph{How does PA frequency affect error accumulation in long-time WBRW simulations?}
    \vspace{0.2cm}
\end{center}

% our contribution
In this work, we answer the above two questions. We first establish an overall error estimation for WBRW-PAUM and make its dependence on PA frequency explicit, which separates branching variance, resampling variance, and squared bias. Our analysis shows that, beyond merely controlling particle growth, PA replaces the exponential-in-time variance bound with a linear-in-time bound, thereby alleviating the numerical sign problem. It further reveals two competing effects of PA frequency on the long-time error: overly frequent annihilation leads to the accumulation of annihilation-induced bias, while insufficient annihilation allows the stochastic variance to grow excessively. These theoretical results are verified by numerical experiments, as also roughly mentioned in \cite{Shao2019,Xiong2019}.
More importantly, building on the newfound error structure, we propose a variance-constrained cost optimization strategy for adaptive PA scheduling. To be more specific, we develop an intrinsic-time reparameterization as a common clock for variance control, PA scheduling, and branching dynamics under time-dependent potentials. The strategy is applicable to PAUM, SPADE, and SPADE-Mix, and numerical experiments show that it lies on the low-cost portion of the Pareto frontier~\cite{Zitzler2000pareto} objecting to balance accuracy and efficiency. 

The rest of this paper is organized as follows. Section~\ref{sec:background} introduces the preliminaries about WBRW and PA. Section~\ref{sec:WBRW} states our theoretical analysis for the effect of PA. Section~\ref{sec:adaptive strategy} presents PA scheduling strategy for general time-dependent potentials. Section~\ref{sec:proof of error estimates} provides the proofs. Section~\ref{sec:as} demonstrates numerical experiments to validate the error estimation, and to assess the performance of our proposed scheduling strategy. Section~\ref{sec:con} concludes the paper and discusses future directions.

\section{Preliminaries}
\label{sec:background}  
We now present a concise description of WBRW \cite{Shao2019,Shao2020} and detail the implementation of PA~\cite{Xiong2024SPADE}. 
\subsection{WBRW}
\label{sec:wbrw process}
WBRW is tailored for the truncated Wigner equation
\begin{equation}
    \label{eq:wigner equation}
    \begin{aligned}
        &\partial_t f(\vb*{x},\vb*{k},t)+\frac{\hbar\vb*{k}}{m}\cdot\nabla_{\vb*{x}}f(\vb*{x},\vb*{k},t)=\Theta_V[f](\vb*{x},\vb*{k},t),\\
        &\Theta_V[f](\vb*{x},\vb*{k},t)=\int_{\mathcal{K}} f(\boldsymbol{x},\boldsymbol{k}',t)V_{W}(\boldsymbol{x},\boldsymbol{k-k'},t)\D\boldsymbol{k'},\\
        &V_W(\vb*{x},\vb*{k},t)=\frac{1}{\imath\hbar(2\pi)^s}\int_{\mathbb{R}^s}\me^{-\imath\vb*{k}\cdot\vb*{y}}(V(\vb*{x}+\frac{\vb*{y}}{2},t)-V(\vb*{x}-\frac{\vb*{y}}{2},t))\D \vb*{y}.
    \end{aligned}
\end{equation}
Here $\vb*{x}$ is the position, $\vb*{k}$ is the wave vector, $m$ is the particle mass, and $\hbar$ is the reduced Planck constant, $V_{W}(\boldsymbol{x},\boldsymbol{k},t)$ is the Wigner kernel in the nonlocal pseudo-differential operator, obtained from the potential $V(\vb*{x},t)$, $f(\boldsymbol{x},\boldsymbol{k},t)$ denotes the Wigner function, $\mathcal{X}\times \mathcal{K}:=\prod_{i=1}^s\left[x_{\min}^{(i)}, x_{\max}^{(i)}\right] \times$ $\prod_{i=1}^s\left[k_{\min }^{(i)}, k_{\max }^{(i)}\right]=:\Omega$ is a bounded and sufficiently large $2s$-dimensional hypercube outside which $f$ diminishes. The Wigner kernel can be decomposed into $M$ positive and negative parts with Hahn-Jordan decomposition, 
\begin{equation}
    \label{eq:multiple kernel}
    \begin{aligned}
        &V_W(\vb*{x},\vb*{k},t)=\sum_{m=1}^M V_{m,W}^+(\vb*{x},\vb*{k},t)-\sum_{m=1}^M V_{m,W}^-(\vb*{x},\vb*{k},t),\\
        &\Theta_{m,V}^{\pm}[f](\vb*{x},\vb*{k},t)=\int_{\mathcal{K}} f(\boldsymbol{x},\boldsymbol{k}',t)V_{m,W}^{\pm}(\boldsymbol{x},\boldsymbol{k-k'},t)\D\boldsymbol{k'}. 
    \end{aligned}
\end{equation}
where $\xi_m(\vb*{x},t) = \int_{2\mathcal{K}} V_{m,W}^{+}\left(\vb*{x}, \vb*{k}, t\right) \D \vb*{k} = \int_{2\mathcal{K}} V_{m,W}^{-}\left(\vb*{x}, \vb*{k}, t\right) \D \vb*{k}$ denotes the normalizing constant of the $m$-th part by its antisymmetry property, and $\xi(\vb*{x},t) = \sum_{m=1}^M \xi_m(\vb*{x},t)$. 

Given $\varphi_T\in L^2(\Omega)$, consider the dual system by $\langle\varphi_T,f(T)\rangle=\langle\varphi(t),f(t)\rangle=\langle\varphi(0),f(0)\rangle$ for any $t\in[0,T]$, i.e., $\frac{\D}{\D t}\langle\varphi(t),f(t)\rangle=0$. Integration by parts and the antisymmetric property of $V_W$ give the adjoint equation with terminal condition \cite{Shao2020}
\begin{equation}
    \label{eq:adjoint wigner equation}
    \begin{cases}
        \partial_t\varphi(\vb*{x},\vb*{k},t)+\frac{\hbar\vb*{k}}{m}\cdot\nabla_{\vb*{x}}\varphi(\vb*{x},\vb*{k},t)
        =\Theta_V[\varphi](\vb*{x},\vb*{k},t),\\
        \varphi(\vb*{x},\vb*{k},T)=\varphi_T(\vb*{x},\vb*{k}).
    \end{cases}
\end{equation}
Subtracting $\gamma_0(s)\varphi$ from both sides of \eqref{eq:adjoint wigner equation}, we obtain the renewal-type integral equation by the method of characteristics and variation of constant, 
\begin{equation}
    \label{eq:wbrw first moment renewal}
    \begin{aligned}
        \varphi(\vb*{x},\vb*{k},t)
          =&\big(1-\mathcal G_t(T-t)\big)\varphi_T(\vb*{x}(T-t),\vb*{k})\\
          &+\int_0^{T-t}\left[\varphi-\frac{\Theta_V[\varphi]}{\gamma_0(t+\tau)}\right](\vb*{x}(\tau),\vb*{k},t+\tau)\,\D\mathcal G_t(\tau),
    \end{aligned}
\end{equation}
whose stochastic interpretation induces the Wigner branching random walk. Here $\vb*{x}(\tau)=\vb*{x}+\hbar\vb*{k}\tau/m$, $\gamma_0(s)$ is positive and integrable on $[0,T]$, and $\mathcal G_t$ is the waiting time distribution for a particle starting at time $t$,
\begin{equation}
    \mathcal G_t(\tau)=1-\exp(-\int_t^{t+\tau}\gamma_0(s)\,\D s).
        \label{eq:exponential clock}
\end{equation}

Iterating Eq.~\eqref{eq:wbrw first moment renewal} gives WBRW. Draw $E\sim\operatorname{Exp}(1)$. If $E\geq\int_t^T\gamma_0(s)\,\D s$, the particle freezes at $(\vb*{x}(T-t),\vb*{k},T)$. Otherwise it branches at $t+\tau$ determined by $\int_t^{t+\tau}\gamma_0(s)\,\D s=E$. If $\xi(\vb*{x}(\tau),t+\tau)>0$, kill the particle, choose component $m$ with probability $\xi_m/\xi$ at that point, draw a single increment $\Delta\vb*{k}$ from $V_{m,W}^+(\vb*{x}(\tau),\cdot,t+\tau)/\xi_m$ and generate three offsprings by
\begin{equation*}
\begin{aligned}
& \boldsymbol{x}_{(1)}=\boldsymbol{x}_{(2)}=\boldsymbol{x}_{(3)}=\vb*{x}(\tau), \quad \vb*{k}_{(1)}=\vb*{k}+\Delta \vb*{k},\quad \vb*{k}_{(2)}=\vb*{k}-\Delta \vb*{k},\quad\boldsymbol{k}_{(3)}=\boldsymbol{k}, \\
& w_{(1)}=w \cdot \frac{\xi(\vb*{x}(\tau),t+\tau)}{\gamma_0(t+\tau)} \mone_{\left\{\boldsymbol{k}_{(1)} \in \mathcal{K}\right\}}, \quad w_{(2)}=-w \cdot \frac{\xi(\vb*{x}(\tau),t+\tau)}{\gamma_0(t+\tau)} \mone_{\left\{\boldsymbol{k}_{(2)} \in \mathcal{K}\right\}}, \quad w_{(3)}=w.
\end{aligned}
\end{equation*}
Here we lower bound $\gamma_0(t)$ to guarantee the boundness of particle weights,
\begin{equation}
    \gamma_0(t)\geq\breve\xi(t):=\sup_{\vb*{x}\in\mathcal X}\xi(\vb*{x},t).
        \label{eq:auxiliary function}
\end{equation}
Given their birth states, the children evolve independently from $t+\tau$. For the $m$-th component, taking expectation of $\varphi$ at $\Delta \vb*{k}$ gives $\Theta_{m,V}^-[\varphi]/\xi_m$ and $\Theta_{m,V}^+[\varphi]/\xi_m$, respectively. Thus for $w=1$ the bracket in Eq.~\eqref{eq:wbrw first moment renewal} equals $\mathbb E[\sum_{j=1}^3 w_{(j)}\varphi(\vb*{x}(\tau),\vb*{k}_{(j)},s)\mid\tau]$. That is, WBRW is naturally an unbiased estimator in the sense that
\begin{equation}
    \label{eq:unbiasedness0}
    \varphi(\vb*{x},\vb*{k},t)
      =\mathbb E\!\left[\langle\varphi_T,\nu_T\rangle
            \mid\nu_t=\delta_{(\vb*{x},\vb*{k})}\right],
       \quad 0\leq t\leq T,
\end{equation}
where $\nu_t$ is the empirical signed measure at time $t$, enjoying a general form of
\begin{equation}
    \label{eq:empirical measure}
    \nu_t=\frac{1}{N_\alpha}\sum_{i=1}^{N(t)}w_i(t)\delta_{(\vb*{x}_i(t),\vb*{k}_i(t))},
\end{equation}
with $N_\alpha$ being the initial particle number, and $N(t)$ the number of particles at time $t$. By the independence of initial particles and the tower property, the unbiasedness extends to any initial distribution $\nu_0$, 
\begin{equation}
    \label{eq:unbiasedness}
    \mathbb E\langle\varphi_T,\nu_T\rangle=\mathbb E\langle\varphi(t),\nu_t\rangle=\mathbb E\langle\varphi(0),\nu_0\rangle.
\end{equation}
In particular, if $\nu_0$ is sampled from the initial Wigner function $f(0)$, then the duality gives the unbiased approximation of the observable $\langle\varphi_T,f(T)\rangle$ by WBRW, 
\begin{equation}
    \langle\varphi_T,f(T)\rangle=\langle\varphi(0),f(0)\rangle=\mathbb E\langle\varphi_T,\nu_T\rangle\approx \frac{1}{N_\alpha}\sum_{i=1}^{N_\alpha}\varphi_T(\vb*{x}_i(T),\vb*{k}_i(T)).
\end{equation}

A PA schedule specifies when annihilation is performed, typically at fixed time intervals. At each scheduled PA time, particles are frozen, and the resampled particles resume evolution with their clocks set to that physical time.

\subsection{PA}
PA is implemented through a two-stage procedure, partitioning the domain and cancelling positive and negative particles. After PA, the expectation of particle number is reduced from $N(t)$ to $\sum_{k=1}^K |W_k|$, where $W_k$ is the sum of weights in the $k$-th bin. It is then straightforward to see the effect of PA on controlling the growth of particle number. In this section, we will illustrate three feasible PA strategies in detail, PAUM, SPADE and SPADE-Mix. 

\subsubsection{PAUM}
\label{sec:PAUM}
PAUM employs the idea of histogram statistics and residual resampling. First, split the whole region $\Omega$ into a uniform mesh. The domain considered generally is a rectangular bin $\Omega$ with dimension $d$. PAUM splits $\Omega=\bigcup_{k=1}^K \mathrm{Q}_k$, where $\mathrm{Q}_k=\mathcal{Y}_{i_1}\times\cdots\mathcal{Y}_{i_d} $ is the tensor product of disjoint rectangular bins, and $K$ denotes the number of bins. For $l=1, \ldots, d$, $\mathcal{Y}_{i_l}=[y_{\text {min }}^{(l)}+\left(i_l-1\right) \Delta y_l, y_{\text {min }}^{(l)}+i_l \Delta y_l]$, and $\Delta y_l$ is state variable stepsize in the $l$-th dimension. The bin functions $\varphi_k=|Q_k|^{-1}\mone_{Q_k}$ give the piecewise constant approximation
\begin{equation}
    \label{eq:piecewise constant approximation}
    \mathcal P_h f(t)\approx\mathcal P_h\nu_t
        =\sum_{k=1}^K |Q_k|\langle\varphi_k,\nu_t\rangle\varphi_k .
\end{equation}

Second, we annihilate particles with opposite signs in the same bin using residual resampling. The resampled particle number $N_k$(t) in $\mathrm{Q}_k$ is the round to the absolute value of the weights sum $W_k$, i.e.,
\begin{equation}
    \label{eq:residual resampling}
    N_k(t) = \begin{cases}
        \lfloor |W_k| \rfloor, & \text{with probability } 1 - \{|W_k|\},\\
        \lfloor |W_k| \rfloor + 1, & \text{with probability } \{|W_k|\} ,
    \end{cases}
\end{equation}
here $\lfloor \cdot \rfloor$ and $\{\cdot\}$ denote the integer and fractional parts of a real number, respectively. Then each resampled particles are regenerated with their state variables $\boldsymbol{y}$ sampled uniformly within their respective bin. Their weights of particles in the bin $Q_k$ are normalized to $+1$ or $-1$ in accord with the sign of the sum of weights $W_k$. 

\subsubsection{SPADE}
\label{sec:SPADE}
As the uniform mesh will be seriously affected by CoD when dealing with high-dimensional problems, it is necessary to introduce an adaptive partition method to achieve PA. SPADE is such a method proposed by \cite{Shao2020SPADE}, overcoming the numerical sign problem encountered in solving high-dimensional problems \cite{Xiong2024SPADE}.

First, the adaptive partition of $\Omega$ is achieved through sequential binary partition. Let $\mathcal{P}$ be a binary splitting on $\Omega$, with initial state $\mathcal{P}_{1}=\{\Omega\}$. When $\Omega$ is partitioned into the $K$ bins, namely, $\Omega=\mathcal{P}_{K}=\bigcup_{k=1}^K \mathrm{Q}_k$,  $\mathcal{P}_{K+1}$ is produced by splitting one of the bins in $\mathcal{P}_{K}$ into two sub-rectangles along one dimension and merging them with the rest of bins in $\mathcal{P}_{K}$.

The criterion for partitioning $\mathrm{Q}_k$ is to control the star discrepancy of positive and negative particles in $\mathrm{Q}_k$, inspired by the ideas from tree-based high-dimensional density estimation~\cite{Li2016}. This converts CoD into the NP-hard problem of calculating the star discrepancy, which can be approximated by external solvers~\cite{gnewuch2012new}. Denote $\mathcal{S}_{k}^{+}:=\left\{\boldsymbol{y}_i^{+} \right\}_{i=1}^{P_{k}(t)}$ and $\mathcal{S}_{k}^{-}:=\left\{\boldsymbol{y}_i^{-}\right\}_{i=1}^{M_{k}(t)}$ as positive and negative particles with weights $\pm 1$ in $\mathrm{Q}_k$ after bootstrap filtering, where $P_k(t)$ and $M_k(t)$ represent the corresponding numbers of positive and negative particles. We need to continue partitioning the bin $\mathrm{Q}_{k}$ until the star discrepancies of positive and negative particles satisfy
\begin{equation}
D_{P_{k}(t)}^{*}\left(\mathcal{S}_{k}^{+}\right) \leq \frac{\vartheta \sqrt{N_{k}(t)}}{\max \left(P_{k}(t), M_{k}(t)\right)} \text{ and } D_{M_{k}(t)}^{*}\left(\mathcal{S}_{k}^{-}\right) \leq \frac{\vartheta \sqrt{N_{k}(t)}}{\max \left(P_{k}(t), M_{k}(t)\right)}, \label{2.6}
\end{equation}
where the parameter $\vartheta$ controls the diameter of the partition and thus adjusts $K$. 

According to the above partition criteria~(\ref{2.6}), if $\mathrm{Q}_k$ needs to be split, then it is necessary to decide along which dimension direction and where to perform the binary partition. We need to select a point $c_{j}^{(k)}\in(a_{j}^{(k)},b_{j}^{(k)})$ in the $j$-th dimension, such that the region $\mathrm{Q}_{k}$ will be split into $\mathrm{Q}_{k}^{(1)}$ and $\mathrm{Q}_{k}^{(2)}$: $\mathrm{Q}_{k}^{(1)}=\prod_{i=1}^{j-1}[a_{i}^{(k)}, b_{i}^{(k)}] \times[a_{j}^{(k)}, c_{j}^{(k)}] \times \prod_{i=j+1}^{d}[a_{i}^{(k)}, b_{i}^{(k)}], \quad \mathrm{Q}_{k}^{(2)}=\mathrm{Q}_{k} \backslash \mathrm{Q}_{k}^{(1)}$. Denote $P_{k}^{(1)}$ and $M_{k}^{(1)}$ as the numbers of positive and negative particles in $\mathrm{Q}_{k}^{(1)}$, respectively. In order to seperate the positive particles and negative particles as much as possible, we need to select a node $c_{j}^{(k)}$ in the $j$-th dimension to optimize the difference gap function from \cite{Shao2020SPADE, Xiong2024SPADE}:
\begin{equation}
g(c_{j}^{(k)}):=\frac{1}{2} \max _{\mathrm{Q}_{k}^{(1)}}\left(\left|\frac{P_{k}^{(1)}}{P_{k}}-\frac{M_{k}^{(1)}}{M_{k}}\right|\right)=\frac{1}{2} \max _{\mathrm{Q}_{k}^{(2)}}\left(\left|\frac{P_{k}^{(2)}}{P_{k}}-\frac{M_{k}^{(2)}}{M_{k}}\right|\right).\label{2.8}
\end{equation}
A practical way to obtain a (sub)-optimal $c_j^{(k)}$ is to let the equidistant points be the candidates, $c_{j, l}^{(k)}=a_j^{(k)}+\frac{l}{m}(b_j^{(k)}-a_j^{(k)})$ in $[a_j^{(k)}, b_j^{(k)}], l=1, \ldots, m-1$. 

When the partition criteria \eqref{2.6} are satisfied for all bins, the partition of the region is finished and random matching is conducted in each bin to cancel out the paired positive and negative particles. 

\subsubsection{SPADE-Mix}
\label{sec:SPADE-Mix}
In SPADE-Mix, we use the mixture discrepancy  instead of the star discrepancy to measure the uniformity of the point set distribution within the bin. Mixture discrepancy admits an explicit analytical formula, enabling efficient and exact computation in $O(L^2 d)$ time for high-dimensional problems. Moreover, the discrepancy preserves desirable properties such as rotational invariance, ensuring robust uniformity measure. More details can be found in \cite{Li2016}. The steps of adaptive bin partition are the same as those of SPADE.

\section{Error estimation}
\label{sec:WBRW}
It is aforementioned that PA controls the growth of the particle number and thus influences the efficiency. Then a direct question arises: \emph{how does PA affect the accuracy and stability?} This section will provide a rigorous error estimation for WBRW-PAUM, answering the question from a theoretical perspective.

We first extend the variance estimate for WBRW-HJD without PA in \cite[Theorem~1]{Shao2020} to time-dependent $\gamma_0(t)$, multi-kernel HJD formulation in Eq.~\eqref{eq:multiple kernel}, and probabilistic selection of kernel components. A sharper bound is obtained. 
\begin{proposition}
    \label{prop:without PA} 
    Let $\Delta\Phi_{w,T}^{(2)}(\vb*{x},\vb*{k},t)=\text{Var}[\langle\varphi_T,\nu_T\rangle\big| \nu_t=\delta_{(\vb*{x},\vb*{k})}]$. Given $\varphi_T\in L^2(\Omega)$, the $L^1$-norm of pointwise variance for WBRW without PA is bounded by
    \begin{equation}
        \label{eq:variance without PA}
        \|\Delta\Phi_{w,T}^{(2)}\|_1\le 2(\me^{\int_t^T\frac{2\breve{\xi}^2(s)}{\gamma_0(s)}\,ds}-1)\|\varphi_T\|_2^2,\quad \forall t\in[0,T],
    \end{equation}
    where $L^1$-norm and $L^2$-norm are defined on $\Omega$. 
\end{proposition}

In the time-independent case, the bound grows exponentially with the evolution time and decreases as the auxiliary branching rate $\gamma_0$ increases, in agreement with \cite[Theorem~1]{Shao2020}. Proposition~\ref{prop:without PA} removes the explicit dependence on the $L^2$-operator bound $K_V$ for $\Theta_V$ from both the exponent and the prefactor, making the effect of $\gamma_0$ more transparent. The exponential growth factor is rooted in the HJD decomposition, in which the positive and negative kernel contributions accumulate without cancellation in the diagonal second moment. Exponential variance growth without PA has also been observed numerically in \cite{Xiong2024SPADE}. 

The proof follows the same steps as in \cite[Theorem~1]{Shao2020}, where the renewal-type integral equation for second moment is derived and the backward Gr\"{o}nwall's inequality is applied to get the exponential variance bound. The key difference is that we employ the antisymmetry of the Wigner kernel when integrating the moment equation, and use $L^2$-conservation for evolution in place of  the original $L^2$-stability estimate. The details of the proof are provided in Section~\ref{sec:proof of error estimates}.

Error estimation with PA requires accounting for the errors induced by annihilation and their propagation through subsequent WBRW evolution. Our analysis combines these effects with WBRW variance estimated above. 
\begin{theorem}
    \label{thm:main theorem}
    Let PA time steps be $t_1<\cdots<t_{R-1}$, and $t_0=0$, $t_R=T$. Then the error estimation for WBRW-PAUM at the terminal time $T$ is given by
    \begingroup
    \clubpenalty=10000
    \setlength{\abovedisplayskip}{4pt plus 1pt minus 1pt}
    \setlength{\belowdisplayskip}{4pt plus 1pt minus 1pt}
    \setlength{\belowdisplayshortskip}{4pt plus 1pt minus 1pt}
    \begin{equation}
        \mathbb{E}\left[\|\mathcal{P}_h\nu_T-f(T)\|_2^2\big| \{t_r\}_{r=0}^{R}\right]\le \mathcal{E}_p+\mathcal{E}_r+\mathcal{E}_w,
    \end{equation}
    where $\mathcal{E}_p$ is the squared bias from piecewise constant projection in PAUM, $\mathcal{E}_w$ is accumulated WBRW variance, and $\mathcal{E}_r$ accounts for initial sampling and residual resampling variance, given by
    \begin{equation}
        \begin{aligned}
            &\mathcal{E}_p = \sum_{r=0}^{R} \|(\mathcal{I}-\mathcal{P}_h)f(t_r)\|_2^2,\\
            &\mathcal{E}_w = \sum_{r=0}^{R-1} 2(\me^{\int_{t_r}^{t_{r+1}}\frac{2\breve{\xi}^2(s)}{\gamma_0(s)}\dd s}-1)\frac{\mathbb{E}[\|\mathcal{P}_h \nu_{t_r}\|_\infty\big| \{t_r\}_{r=0}^{R}]|\Omega|}{N_\alpha|Q|},\label{eq:variance WBRW}\\
            &\mathcal{E}_r = \mathbb E\left[\|\mathcal P_h\nu_0-\mathcal P_h f(0)\|_2^2\right]+\sum_{r=0}^{R-1}\left(\frac{\mathbb E[\|\mathcal P_h\nu_{t_r}\|_\infty\big|\{t_r\}_{r=0}^{R}]|\Omega|}{N_\alpha|Q|}+\frac{|\Omega|}{4N_\alpha^2|Q|^2}\right),
        \end{aligned}
    \end{equation}
    Furthermore, if $f\in H^1(\Omega)$, then the projection error can be bounded by Poincaré inequality \cite{Payne1960,Stanoyevitch1993poincare}
    \begin{equation}
        \label{eq:projection error}
        \mathcal{E}_p \le \sum_{r=0}^{R}\left( (\frac{h_{\vb*{x}}}{\pi})^2\|\nabla_{\vb*{x}} f(t_r)\|_2^2 + (\frac{h_{\vb*{k}}}{\pi})^2\|\nabla_{\vb*{k}} f(t_r)\|_2^2\right),
    \end{equation}
    \endgroup
    where $h_{\vb*{x}}$ and $h_{\vb*{k}}$ are the mesh sizes in $\vb*{x}$ and $\vb*{k}$ directions, respectively.
\end{theorem}

Theorem~\ref{thm:main theorem} decomposes the overall error into three parts: WBRW variance $\mathcal{E}_w$, sampling and resampling variance $\mathcal{E}_r$, and projection squared bias $\mathcal{E}_p$. $\mathcal{E}_w$ grows exponentially with time between consecutive PA steps, while $\mathcal{E}_r$ increases with each resampling step. For fixed mesh and PA schedule, $\mathcal{E}_w$ and $\mathcal{E}_r$ have leading order $N_\alpha^{-1}$ under independent unbiased initialization, giving half-order convergence of the root mean squared $L^2$ sampling error. $\mathcal{E}_p$ is $\mathcal{O}(h_{\vb*{x}}^2+h_{\vb*{k}}^2)$ with respect to the mesh sizes $h_{\vb*{x}}$ and $h_{\vb*{k}}$ according to Poincaré inequality, and accumulates over time steps. There is no time discretization error in bias, since we introduce the spatial-independent auxiliary function $\gamma_0(t)$ to transform the original PDE into a renewal-type integral equation and simulate it by a continuous-time branching random walk, as \cite{Shao2019} has discussed.

More importantly, Theorem~\ref{thm:main theorem} demonstrates the effect of PA on the variance control and long-time stability. Define $\Lambda(t):=\int_0^t 2\breve{\xi}^2(s)/\gamma_0(s)\dd s$, and let $x_r:=\Lambda(t_{r+1})-\Lambda(t_r)$, $x_{\max}=\max_{0\le r\le R-1}x_r$. Then we have
\begin{equation}
    \label{eq:variance bound}
    \mathcal{E}_w
    \le \frac{2C}{N_\alpha}\sum_{r=0}^{R-1}(\me^{x_r}-1)
    \le \frac{2C}{N_\alpha}\mathcal{V}(x_{\max})\Lambda(T),
\end{equation}
where $C=\max_{0\le r\le R-1}\mathbb{E}[\|\mathcal{P}_h \nu_{t_r}\|_\infty | \{t_r\}_{r=0}^{R}]|\Omega|/|Q|$ and $\mathcal{V}$ is strictly increasing and defined by $\mathcal{V}(x):=\frac{\me^x-1}{x}$, $x>0$. WBRW without PA is a special case of WBRW-PAUM with $R=1$ and $t_1=T$, where the variance bound grows exponentially as $\mathcal{E}_w\le 2C(\me^{\Lambda(T)}-1)/N_\alpha$. \emph{Hence introducing PA changes the final time dependence of WBRW variance bound from $\me^{\Lambda(T)}-1$ to $\Lambda(T)$, provided that $x_{\max}$ and $C$ are uniformly bounded in $T$.} This parallels the study of how error constants depend on the final time $T$ in classical long-time numerical analysis. Given a prescribed WBRW variance tolerance $\varepsilon$ and fixed $N_\alpha$, the time allowed by the corresponding bounds are
\begin{equation}
    \Lambda(T)\le
    \ln\left(1+\frac{N_\alpha\varepsilon}{2C}\right)
    \quad\text{without PA},
    \quad
    \Lambda(T)\le
    \frac{N_\alpha\varepsilon}{2C\mathcal{V}(x_{\max})}
    \quad\text{with PA}.
\end{equation}
Thus PA allows the long-time simulation of WBRW by controlling the variance growth, highlighting its importance in long-time stability. 

However, choosing the PA frequency is challenging because it requires balancing competing effects on accuracy and computational cost. On one hand, frequent PA controls WBRW variance $\mathcal{E}_w$ and limits particle growth between PA steps,   
\begin{equation}
    \label{eq:parnum}
    \mathbb{E}[N(t_{r+1})|N(t_r)]
    \le\me^{\int_{t_r}^{t_{r+1}}2\gamma_0(s)\dd s}
    \mathbb{E}N(t_r),
    \quad t\in[t_r,t_{r+1}),
\end{equation}
which is stated in \cite[Theorem 5.5]{Shao2019}, thereby reducing the cost of subsequent PA steps. On the other hand, each PA step introduces projection bias and resampling variance to $\mathcal{E}_p$ and $\mathcal{E}_r$, respectively, and incurs partitioning overhead, particularly for SPADE and SPADE-Mix. Excessively frequent PA can therefore increase accumulated PA error and computational overhead.

\begin{figure}[htbp]
    \centering
    \includegraphics[width=.75\textwidth]{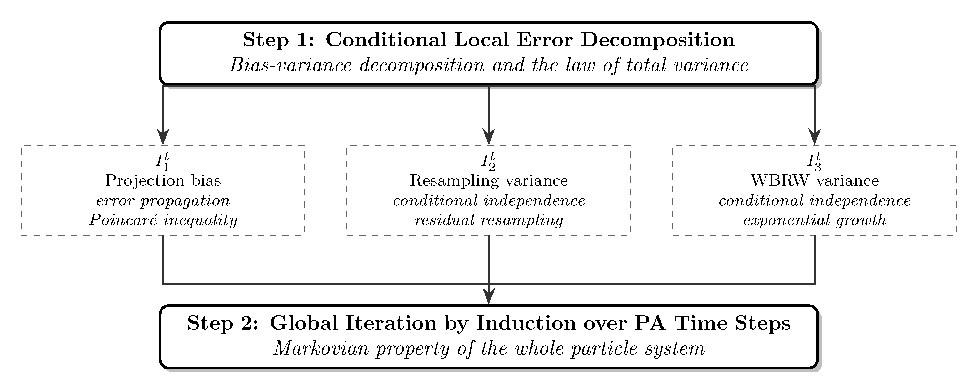}
    \caption{The proof of Theorem~\ref{thm:main theorem}: a flowchart to illustrate the error decomposition, seen in Eq.~\eqref{eq:weak error decomposition}, and propagation. Full notations and proof details can be found in Section~\ref{sec:proof of error estimates}.}
    \label{fig:flowchart}
\end{figure}

Finally, we would like to summarize the proof structure of Theorem~\ref{thm:main theorem} in Figure~\ref{fig:flowchart}. The bias-variance decomposition and the law of total variance are powerful tools to decompose the overall error into three parts. Besides, two fundamental properties play a crucial role: the Markovian property of the whole system allowing us to propagate the error, and the conditional independence of each particle making it possible to decouple the overall variance. In fact, the proof structure is a general framework for particle methods with PA~\cite{huang2026AAMM,lei2025JCP}. Complete proof can be found in Section~\ref{sec:proof of error estimates}.
\begin{remark}
    \label{rmk:analogy}
    In analogy with Theorem~\ref{thm:main theorem}, we employ the intuition that the overall error of SPADE and SPADE-Mix can also be decomposed into three main parts: PA bias, resampling variance and WBRW variance. Importantly, $\mathcal{E}_w$ serves as a good estimator of WBRW variance for all PA methods, as verified numerically in Figure~\ref{fig:benchmark-comparison}.
\end{remark}

\section{Variance-constrained cost optimization}
\label{sec:adaptive strategy}
To improve the efficiency under acceptable accuracy, we propose a variance-constrained cost optimization strategy to determine the key parameters: PA time step $\tau$ and branching rate $\gamma_0$. 

Three key ideas are employed. First, it is better to describe the acceptable accuracy from the perspective of controlled variance rather than total error, because the latter is difficult to quantify due to the unknown PA bias mechanism. Second, since the total variance and computational cost accumulate over successive PA steps, a per-unit-time formulation, rather than per-step, is more relevant to the overall accuracy and efficiency. Third, it is useful to measure the interval length in the intrinsic scales governing variance and cost, and to relate these scales through a constant ratio. 

\subsection{Time scales and reparameterization}
\label{subsec:choice variance}
We first expose the scale relations. For a PA interval $[t,t+\tau]$, define
\begin{equation}
\label{eq:choice time scales}
 \begin{aligned}
 x(\tau;t)=\int_t^{t+\tau}\frac{2\breve\xi^2(s)}{\gamma(s)}\dd s,\quad
 y(\tau;t)=\int_t^{t+\tau}2\breve\xi(s)\dd s,\quad
 z(\tau;t)=\int_t^{t+\tau}2\gamma(s)\dd s.
 \end{aligned}
\end{equation}
The three variables characterize three exponential growth scales over a PA interval. Specifically, $x$ controls the WBRW variance bound in Eq.~\eqref{eq:variance WBRW}, while $z$ controls the particle number bound as Eq.~\eqref{eq:parnum}, representing the scale of WBRW or PAUM cost. Lemma~\ref{lem:choice weight} below shows that $y$ controls the growth of the total absolute weight.
\begin{lemma}
    \label{lem:choice weight}
    Let $W(t)=\sum_{i=1}^{N(t)}|w_i(t)|$ be the total absolute particle weight at time $t$. Then, for any PA interval $[t,t+\tau]$, we have
    \begin{equation}
        \label{eq:choice weight}
        \mathbb{E}[W(t+\tau)\big|W(t)] \le W(t)\me^{y(\tau;t)}.
    \end{equation}
\end{lemma}
\begin{proof}[Proof of Lemma~\ref{lem:choice weight}]
    Conditioning on the first branching time and using the offspring weights of Section~\ref{sec:wbrw process} gives $\mathbb{E}[W(t+\tau)\big|W(t)]\le (1-G_t(\tau))W(t)+\int_t^{t+\tau}(1+2\breve{\xi}(s)/\gamma_0(s))\mathbb{E}[W(s)\big|W(t)]\dd \mathcal{G}_t(s-t)$. The Gr\"onwall's inequality gives the bound. 
\end{proof}
Note that boostrap produces particles of absolute weight $1$, and preserves the total absolute weight. It reflects that $y$ is the scale for the particle number after bootstrap. Previous numerical studies \cite{Lei2025discrepancy,Xiong2019} have shown that the costs of SPADE and SPADE-Mix scale approximately linearly with the particle number after bootstrap, so $y$ also indicates the scale of SPADE-like cost. The two cost scales lead to an interpolating model. After removing a positive common coefficient, the computational cost of one PA interval can be modelled by
\begin{equation}
    \label{eq:mixed cost model}
    c(y,z)=\eta e^y+(1-\eta)e^z,
\end{equation}
where $0\leq\eta\leq1$ relies on the specific implementation of PA methods. 

For fixed $x$ and $y$, the variance contribution and the SPADE-like cost scale are fixed, while the WBRW/PAUM cost $\me^z$ is increasing with $z$. By the Cauchy-Schwarz inequality, $xz\ge y^2$, so the smallest admissible WBRW cost scale is $z=y^2/x$. Equality holds if and only if $\gamma_0(s)/\breve{\xi}(s)$ is constant almost everywhere on the interval. This motivates the proportional choice with constant ratio $\rho\ge 1$. That is, all three growth scales are parameterized by the single scale $y$, while $\rho$ controls their relative balance, 
\begin{equation}
\label{eq:choice scale relations}
 x=\frac y\rho,\quad z=\rho y, \quad \gamma_0(t)=\rho\breve{\xi}(t).
\end{equation}
It identifies $y$ as a natural intrinsic time scale and motivates time reparameterization
\begin{equation}
\label{eq:intrinsic theta}
\theta(t)=\int_0^t2\breve{\xi}(s)\mathrm ds,\quad y(\tau;t)=\theta(t+\tau)-\theta(t).
\end{equation} 

\subsection{Optimal PA interval and branching rate}
\label{subsec:choice mixed}
Our optimization strategy is to minimize the cost under a variance constraint, where both cost and variance are measured with per-unit-$y$ formulation, instead of per-step. 

To control the variance, we recall the bound in Eq.~\eqref{eq:variance bound} and prescribe a tolerance $n_V>1$ on the per-unit-$x$ variance, which gives the admissible interval by
\begin{equation}
\label{eq:variance constraint}
 \mathcal V(x)=\frac{\me^x-1}{x}\leq n_V \quad \Leftrightarrow\quad x\leq x_V \quad \Leftrightarrow \quad y\le \rho x_V,
\end{equation}
where $x_V>0$ is uniquely determined by $\mathcal V(x_V)=n_V$, guaranteed by the monotonicity of $\mathcal V$ and $\mathcal{V}(0)=1$. For a partition with local scales $x_r\leq x_V$, the accumulated variance over the entire simulation is then bounded by constant multiples of 
\begin{equation*}
    \sum_{r=0}^{R-1} (\me^{x_r}-1)\leq \sum_{r=0}^{R-1} n_V x_r=n_V\int_0^T \frac{2\breve{\xi}^2(s)}{\gamma(s)}\dd s,
\end{equation*}
which means the admissible intervals controls the total variance by $n_V$ times the baseline of sufficiently small PA intervals. 

Now we turn to minimize the computational cost with known $n_V$ and $x_V$. We have introduced the mixed cost model in Eq.~\eqref{eq:mixed cost model}. The per-unit-$y$ cost with the above variance constraint leads to the following dimensionless optimization problem
\begin{equation}
\label{eq:choice mixed cost}
 \begin{cases}
    \min_{y,\rho}&  \mathcal C(y,\rho)=\frac{c(y,\rho y)}{y}=\frac{\eta e^y+(1-\eta)e^{\rho y}}{y},\\
    \text{s.t.}&  0<y\leq\rho x_V,\quad\rho\geq1.\\
 \end{cases}
\end{equation}
The minimizer of this problem can be given explicitly by Proposition~\ref{prop:choice optimum} below.
\begin{proposition}
\label{prop:choice optimum}
An optimal branching ratio for \eqref{eq:choice mixed cost} is
\begin{equation}
\label{eq:choice optimal rho}
    y^*=\begin{cases}
    \rho^*x_V,&0<x_V<1,\\
    1,&x_V\geq1,
    \end{cases}\qquad 
 \rho^*=\begin{cases}
 \rho_\Phi,&0<x_V<(2-\eta)^{-1},\\
 1,&x_V\geq(2-\eta)^{-1},
 \end{cases}
\end{equation}
where $\rho_\Phi$ is the unique solution to $\Phi(\rho)=\eta e^{\rho x_V}(\rho x_V-1)
 +(1-\eta)e^{\rho^2x_V}(2\rho^2x_V-1)$, 
and can be found by bisection on $\left[\max\{1,(2x_V)^{-1/2}\},\ x_V^{-1}\right]$ when needed.
\end{proposition}

The two thresholds have distinct roles. The value $x_V=1$ determines whether variance limits the optimal interval; $x_V=(2-\eta)^{-1}$ determines whether increasing the branching ratio reduces cost. The proposition also indicates that $\rho^*=y^*=1$ is often a good choice with variance tolerance $n_V\geq e-1$, which will be verified in Section~\ref{sec:as}. SPADE and PAUM are recovered as special cases of Proposition~\ref{prop:choice optimum} by setting $\eta=1$ and $\eta=0$, respectively. 
\begin{corollary}[SPADE]
\label{subsec:choice endpoints}
\label{cor:choice spade}
For $\eta=1$, the optimal parameters are
\begin{equation}
\label{eq:choice spade}
 y^*=1,\qquad \rho^*=\max\{1,x_V^{-1}\}.
\end{equation}
\end{corollary}

\begin{corollary}[PAUM]
\label{cor:choice paum}
For $\eta=0$, the optimal parameters are
\begin{equation}
\label{eq:choice paum rho}
 y^*=\min\{\rho^*x_V,1/\rho^*\}=\begin{cases}
    \sqrt{x_V/2},&0<x_V<1/2,\\
    x_V,&1/2\leq x_V<1,\\
    1,&x_V\geq1,
 \end{cases}
 \qquad \rho^*=\max\{1,(2x_V)^{-1/2}\}.
\end{equation}
\end{corollary}

With the optimal parameters (computed by Algorithm~\ref{optimal alg}) at our disposal, we can show that the modeled cost, WBRW variance bound and accumulated projection bias scale linearly with the intrinsic horizon $\theta(T)$, given that the corresponding coefficients are uniformly bounded. Note that Proposition~\ref{prop:choice optimum} and both corollaries determine the constant $\rho^*$ and $y^*$. Over the fixed intrinsic horizon $\theta(T)$, a PA schedule with uniform interval $y$ approximately performs PA $\theta(T)/y$ times, giving the total-cost model $\mathcal C_{\mathrm{total}}\approx\theta(T)\mathcal C(y,\rho)$. Since $\theta(T)$ is independent of the selected parameters, minimizing $\mathcal{C}(y,\rho)$ approximates to minimize the total cost. According to Eq.~\eqref{eq:variance bound}, the accumulated WBRW variance satisfies $\mathcal E_w\le 2C n_V\theta(T)/(N_\alpha\rho^*)$, and by Eq.~\eqref{eq:projection error} the accumulated squared projection bias is controlled by $b^2\theta(T)/y^*$ up to endpoint terms, where $b^2$ is the upper bound of squared bias per PA step. 

The optimal physical interval can be recovered from
\begin{equation}
\label{eq:choice inverse time}
 \tau^*(t)=\theta^{-1}(\theta(t)+y^*)-t,
\end{equation}
whenever the target $t+\tau^*$ lies within the horizon $[0,T]$. Uniform spacing in intrinsic time therefore produces adaptive spacing in physical time: PA intervals shorten when $\breve{\xi}(t)$ is large and lengthen when $\breve{\xi}(t)$ is small.

The intrinsic-time coordinate also simplifies the sampling of branching process. Starting from physical time $t$, the next branching event at $t+\tau_b$ is determined by $\int_t^{t+\tau_b}\gamma^*(s)\,\mathrm ds=E\sim\operatorname{Exp}(1)$. Since $\gamma^*(s)=\rho^*\breve{\xi}(s)$, this is equivalent to
\begin{equation}
    \theta(t+\tau_b)-\theta(t)=\frac{2E}{\rho^*} \implies \tau_b=\theta^{-1}\left(\theta(t)+\frac{2E}{\rho^*}\right)-t.
\end{equation}
Hence branching becomes a homogeneous Poisson process with rate $\rho^*/2$ in intrinsic time, while consecutive PA steps are separated by the fixed interval $y^*$. The intrinsic-time coordinate therefore provides a common clock for variance control, PA scheduling, and branching dynamics. 

\subsection{Calibration and pratical implementation}
\label{subsec:choice calibration}
Given two dimensionless inputs, $n_V$ for the variance tolerance and $\eta$ for the relative cost of SPADE and WBRW, Algorithm~\ref{optimal alg} determines the entire PA schedule. Here we discuss the calibration of these inputs and several practical implementations.

\begin{algorithm}[htbp]
\caption{Variance-constrained selection in intrinsic time}
\label{optimal alg}
\hspace*{0.02in}{\bf Input:} Tolerance $n_V>1$, cost weight $0\leq\eta\leq1$, and envelope $\breve\xi(s)>0$ on $[0,T]$.\\
\hspace*{0.02in}{\bf Output:} $x_V$, optimal $\rho^*,y^*$, branching rate $\gamma^*(s)$, and physical PA interval $\tau^*(t)$.
\begin{algorithmic}[1]
\State Set $u\gets1$; double $u$ until $(e^u-1)/u\geq n_V$.
\State Bisect $(e^x-1)/x=n_V$ on $[0,u]$ for $x_V$, using value $1$ at $x=0$.
\If{$\eta=1$ (SPADE)}
    \State $\rho^*\gets\max\{1,y_{\text{FH}}/x_V\},\quad y_{\text{FH}}\gets \theta(T)/(\lceil\theta(T)\rceil-1)$ \hfill [\eqref{eq:choice spade} and \eqref{eq:choice finite horizon spade}]
\ElsIf{$\eta=0$ (PAUM)}
    \State $\rho^*\gets\max\{1,(2x_V)^{-1/2}\}$ \hfill [\eqref{eq:choice paum rho}].
\ElsIf{$x_V\geq(2-\eta)^{-1}$}
    \State $\rho^*\gets1$.
\Else
    \State Bisect on $[\max\{1,(2x_V)^{-1/2}\},x_V^{-1}]$ for $\rho^*$:
    \Statex \hspace{\algorithmicindent}\hspace{\algorithmicindent}\hspace{\algorithmicindent}\hspace{\algorithmicindent}$\eta e^{\rho x_V}(\rho x_V-1)+(1-\eta)e^{\rho^2x_V}(2\rho^2x_V-1)=0$.
\EndIf
\State $y^*\gets\min\{\rho^*x_V,1\}$ and $\gamma^*(s)\gets\rho^*\breve\xi(s)$.
\State Construct $\theta(t)=\int_0^t2\breve\xi(s)\dd s$ and its monotone inverse.
\State $\tau^*(t)\gets\theta^{-1}(\min\{\theta(t)+y^*,\theta(T)\})-t$.
\State \Return $(x_V,\rho^*,y^*,\gamma^*,\tau^*)$.
\end{algorithmic}
\end{algorithm}

\paragraph{Variance tolerance}
PA bias guides the choice of $n_V$. For example, let $b^2$ be the squared bias per PA step and consider the local error model $[C_V(e^x-1)+C_Bb^2]/x$, with positive, locally fixed coefficients $C_V\propto N_\alpha^{-1}$ and $C_B$. Its stationary condition is
\begin{equation}
\label{eq:choice bias balance}
 e^x(x-1)+1=\frac{C_Bb^2}{C_V}.
\end{equation}
The left-hand side increases from zero and determines a variance scale $x$, hence a tolerance $n_V=\mathcal V(x)$. For PAUM, $b^2=O(h^2)$, so the calibration depends on $N_\alpha h^2$: halving the mesh size and quadrupling the initial particle number preserves this balance. Lower PA bias supports a smaller tolerance and more frequent annihilation. Section~\ref{sec:as} will examine the choice of $n_V$ for the three PA methods.

\paragraph{Cost weight}
$\eta$ can be estimated by fitting CPU cost per PA cycle to $Ae^y+Be^z$, with $A,B\geq0$, and setting $\eta=A/(A+B)$. Here $\eta=0.9$ as a default for SPADE-Mix. 

\paragraph{From intrinsic time to physical time}
In practice, the cumulative map $\theta$ in Eq.~\eqref{eq:intrinsic theta} can be evaluated by numerical quadrature and inverted by linear interpolation. If $\breve\xi(t)\ge m>0$ and the intrinsic-time residual is bounded by $\varepsilon$, then $|\widehat t-t|\le \frac{\varepsilon}{2m}$. Thus the inversion introduces only a controllable numerical error, which can be made negligible compared with the particle and annihilation errors. If $\breve\xi$ vanishes on an interval, then $\theta$ is constant and $\gamma^*=\rho^*\breve\xi$ also vanishes there, so no branching occurs on that interval. A generalized inverse of $\theta$ may then be used.

\paragraph{Finite horizon}
Over a finite horizon $T$, the number of PA events is discrete. If PA is performed at intrinsic times $y,2y,\ldots$ strictly before $\theta(T)$, then the number of PA events is $M(T,y)=\lceil\theta(T)/y\rceil-1$. Hence a small increase in $y$ can sometimes remove one PA step, which can be beneficial to improve the efficiency for PA method with high cost. For SPADE, the continuous optimum is $y^*=1$. When $\theta(T)>1$, the smallest such adjustment is
\begin{equation}
\label{eq:choice finite horizon spade}
    y_{\mathrm{FH}}
    =
    \frac{\theta(T)}{\lceil\theta(T)\rceil-1},
    \quad
    \rho_{\mathrm{FH}}
    =
    \max\left\{1,\frac{y_{\mathrm{FH}}}{x_V}\right\}.
\end{equation}
Let $\delta=y_{\mathrm{FH}}-1$. In the SPADE cost model $\mathcal C$, the relative increase from the continuous optimum is $\mathcal C(y_{\mathrm{FH}},\rho_{\mathrm{FH}})/\mathcal C(1,\rho^\ast)-1=e^\delta/(1+\delta)-1=\delta^2/2+O(\delta^3)$. Since $\delta=O(\theta(T)^{-1})$, the finite-horizon adjustment saves one SPADE call relative to $y^\ast=1$ while increasing the modeled cost by only $O(\theta(T)^{-2})$ in relative terms as $\theta(T)\to\infty$.

\section{Proofs}
\label{sec:proof of error estimates}
We first provide a rigorous proof of Proposition~\ref{prop:without PA} and then use the resulting WBRW variance estimate to establish Theorem~\ref{thm:main theorem} for WBRW-PAUM. Finally, we prove Proposition~\ref{prop:choice optimum}, which gives the explicit solution of the optimization problem used in the PA scheduling.

\subsection{Variance estimate without PA}
We extend the proof of \cite[Theorem~1]{Shao2020} and obtain a sharper variance bound for WBRW without PA. 

\begin{lemma}[$L^2$-conservation]
    \label{lem:L^2 conservation}
    Let $\varphi$ be the solution of adjoint Wigner equation Eq.~\eqref{eq:adjoint wigner equation}. Then $\|\varphi(t)\|_2=\|\varphi_T\|_2$ for any $t\in[0,T]$. 
\end{lemma}
\begin{proof}[Proof of Lemma~\ref{lem:L^2 conservation}]
     Integrating the adjoint Wigner equation Eq. \eqref{eq:adjoint wigner equation} with respect to $(\vb*{x},\vb*{k})$ after multiplying $\varphi(\vb*{x},\vb*{k},t)$, we have
    \begin{equation*}
        \frac{1}{2}\frac{\D}{\D t}\|\varphi(t)\|_2^2 + \langle \varphi, \frac{\hbar\vb*{k}}{m}\cdot\nabla_{\vb*{x}}\varphi\rangle = \langle \varphi, \int_{\mathcal{K}} \varphi(\vb*{x},\vb*{k'},t)V_W(\vb*{x},\vb*{k-k'},t)\D \vb*{k'}\rangle.
    \end{equation*}
    Since the support of $\varphi$ is subset of $\Omega$, the second term on the left hand side vanishes using integration by parts. By the antisymmetry of $V_W$, the right hand side also diminishes. Therefore, we have $\frac{\D}{\D t}\|\varphi(t)\|_2^2=0$, which implies the $L^2$ conservation. 
\end{proof}

\begin{lemma}[backward Gr\"{o}nwall's inequality]
    \label{lem:backward gronwall}
    Let $\gamma_0,\beta\in L^1(0,T)$ satisfy $\gamma_0>0$ and $\gamma_0+\beta\ge0$ almost everywhere. Suppose $u$ satisfies the inequality $u(t)\le \alpha(t)+\int_t^{T}(1+\beta(s)/\gamma_0(s))u(s)\,\D\mathcal G_t(s-t)$. Then
    \begin{equation}
        u(t)\le \alpha(t)+\int_t^T(\gamma_0(s)+\beta(s))\me^{\int_t^s\beta(\tau)\,\D\tau} \alpha(s)\,\D s.
    \end{equation}
\end{lemma}
\begin{proof}[Proof of Lemma~\ref{lem:backward gronwall}]
    Let $\tilde{u}(t)=\me^{-\int_0^t\gamma_0(s)\,\D s}u(t)$, $\tilde{\alpha}(t)=\me^{-\int_0^t\gamma_0(s)\,\D s}\alpha(t)$. Then $\tilde{u}(t)\le \tilde{\alpha}(t)+\int_t^{T}(\gamma_0(s)+\beta(s))\tilde{u}(s)\,\D s$. By Gronwall's inequality, we have $\tilde{u}(t)\le \tilde{\alpha}(t)+\int_t^T(\gamma_0(s)+\beta(s))\me^{\int_t^s(\gamma_0(\tau)+\beta(\tau))\,\D\tau} \tilde{\alpha}(s)\,\D s$. Subtituting $\tilde{u}$ and $\tilde{\alpha}$ gives the result.
\end{proof}

\begin{proof}[Proof of Proposition~\ref{prop:without PA}]
Denote the second moment for WBRW starting from a single particle at $(\vb*{x},\vb*{k},t)$ by
\begin{equation}
    \Phi_{w,T}^{(2)}(\vb*{x},\vb*{k},t)
            =\mathbb{E}\left[\langle\varphi_T,\nu_T\rangle^2\mid\nu_t=\delta_{(\vb*{x},\vb*{k})}\right].
\end{equation}
Then the variance satisfies $\Delta\Phi_{w,T}^{(2)}=\Phi_{w,T}^{(2)}-\varphi^2$, where $\varphi$ is defined as the conditional expectation in Eq.~\eqref{eq:unbiasedness} and follows the adjoint Wigner equation in Eq.~\eqref{eq:adjoint wigner equation}. Similar to the proof of \cite[Theorem~1]{Shao2020}, we decompose the second moment according to whether the first branching occurs in $[t,T]$ and which kernel component is selected. In the square of the terminal sum $\Phi_{w,T}^{(2)}(\vb*{x},\vb*{k},t)$, childs contribute $w_{(j)}^2\Phi_{w,T}^{(2)}(\vb*{x}(\tau),\vb*{k}_{(j)},t+\tau)$, and related cross terms $w_{(i)}w_{(j)}\varphi(\vb*{x}(\tau),\vb*{k}_{(i)},t+\tau)\varphi(\vb*{x}(\tau),\vb*{k}_{(j)},t+\tau)$. The resulting renewal-type integral equation for the second moment is then given by
\begin{equation}
    \label{eq:wbrw second moment renewal}
    \begin{aligned}
        \Phi_{w,T}^{(2)}(\vb*{x},\vb*{k},t)
          ={}&\big(1-\mathcal G_t(T-t)\big)\varphi_T^2(\vb*{x}(T-t),\vb*{k})\\
             &+\int_0^{T-t}
                \big(B_w[\Phi_{w,T}^{(2)}]+C\big)(\vb*{x}(\tau),\vb*{k},t+\tau)
                    \,\D\mathcal G_t(\tau).
    \end{aligned}
\end{equation}
At $(\vb*{x},\vb*{k},t)$, we obtain the same diagonal and correlated components, with $\mathbb{E}_m$ denoting the expectation over the $m$-th kernel component, 
\begin{equation*}
    \begin{aligned}
        &B_w[\Phi_{w,T}^{(2)}](\vb*{x},\vb*{k},t)
          =\Phi_{w,T}^{(2)}(\vb*{x},\vb*{k},t)+\frac{\xi(\vb*{x},t)}{\gamma_0^2(t)}\sum_{m=1}^M\big(\Theta_{m,V}^+[\Phi_{w,T}^{(2)}]+\Theta_{m,V}^-[\Phi_{w,T}^{(2)}]\big)(\vb*{x},\vb*{k},t),\\
        &C(\vb*{x},\vb*{k},t)=-\frac{2\varphi\Theta_V[\varphi](\vb*{x},\vb*{k},t)}{\gamma_0(t)}-\frac{2\xi^2(\vb*{x},t)}{\gamma_0^2(t)}\sum_{m=1}^M\frac{\xi_m(\vb*{x},t)}{\xi(\vb*{x},t)}\mathbb{E}_m[\varphi(\vb*{x},\vb*{k}_{(1)},t)\varphi(\vb*{x},\vb*{k}_{(2)},t)].
    \end{aligned}
\end{equation*}

By the triangular inequality and $\|\Theta_{m,V}^\pm[\Phi_{w,T}^{(2)}](\vb*{x},\cdot,t)\|_{L_{\vb*{k}}^1}\le \xi_m(\vb*{x},t)\|\Phi_{w,T}^{(2)}(t)\|_{L_{\vb*{k}}^1}$, we can readily verify the bound $\|B_w[\Phi_{w,T}^{(2)}](t)\|_1\leq (1+2\breve\xi^2(t)/\gamma_0^2(t))\|\Phi_{w,T}^{(2)}(t)\|_1$. Also, the antysymmetry of Wigner kernel gives $\int_{\mathcal K}\varphi\,\Theta_V[\varphi]\,\D\vb*{k}=0$ at each $(\vb*{x},s)$. Thus the first term of $C(\vb*{x},\vb*{k},t)$ diminishes under integration over $\Omega$. For the second term, we use $-2ab\le a^2+b^2$ and then obtain $-2\xi_m\mathbb{E}_m[\varphi(x,k_{(1)},t)\varphi(x,k_{(2)},t)]\le \Theta_{m,V}^+[\varphi^2](\vb*{x},\vb*{k},t)+\Theta_{m,V}^-[\varphi^2](\vb*{x},\vb*{k},t)$. Employing the same argument as for $B_w$ and $L^2$ conservation of $\varphi$ in Lemma~\ref{lem:L^2 conservation} after taking the $L^1$ norm, we have $\int_\Omega C(\vb*{x},\vb*{k},t)\,\D\vb*{x}\,\D\vb*{k}\le 2\breve\xi^2(t)/\gamma_0^2(t)\|\varphi_T\|_2^2$.

Since $\Phi_{w,T}^{(2)}(t)$ is nonnegative, its $L^1$ norm is just the integral of Eq.~\eqref{eq:wbrw second moment renewal} over $\Omega$. Therefore, we can derive the following $L^1$ bound for the second moment, 
\begin{equation}
    \begin{aligned}
        \frac{\|\Phi_{w,T}^{(2)}(t)\|_1}{\|\varphi_T\|_2^2} \le& 1-\mathcal{G}_t(T-t) + \int_t^{T}\frac{2\breve\xi^2(s)}{\gamma_0^2(s)}\D\mathcal G_t(s-t)\\
        &+\int_t^{T}\left[(1+\frac{2\breve\xi^2(s)}{\gamma_0^2(s)})\frac{\|\Phi_{w,T}^{(2)}(s)\|_1}{\|\varphi_T\|_2^2}\right]\D\mathcal G_t(s-t).
    \end{aligned}
\end{equation}
Take $u(t)=\|\Phi_{w,T}^{(2)}(t)\|_1/\|\varphi_T\|_2^2$, $\alpha(t)=1-\mathcal{G}_t(T-t)+\int_t^{T}\frac{2\breve\xi^2(s)}{\gamma_0^2(s)}\D\mathcal G_t(s-t)$ and $\beta(t)=2\breve\xi^2(t)/\gamma_0(t)$. Then we can apply Lemma~\ref{lem:backward gronwall} to obtain
\begin{equation}
    \|\Delta\Phi_{w,T}^{(2)}(t)\|_1=\|\Phi_{w,T}^{(2)}(t)\|_1-\|\varphi(t)\|_2^2\le 2(\me^{\int_t^T2\breve{\xi}^2(s)/\gamma_0(s)\,\D s}-1)\|\varphi_T\|_2^2.
\end{equation}
\end{proof}

\subsection{Error estimate for WBRW-PAUM}

% basis preparation
Recall the basis functions $\{\varphi_k\}_{k=1}^K$ and the piecewise constant projection operator $\mathcal{P}_h$ defined in Eq.~\eqref{eq:piecewise constant approximation}. The projection identity is then derived by the orthogonality of $\{\varphi_k\}_{k=1}^K$, 
\begin{equation}
    \label{eq:projection identity}
    \| \mathcal{P}_h f\|_2^2 = \sum_k |Q_k| \langle f, \varphi_k \rangle^2\le \sum_k |Q_k| \langle f^2, \varphi_k \rangle = \|f\|_2^2. 
\end{equation}

% particle system preparation
We also revisit the particle system for WBRW-PAUM. Denote $t_1>t_0$ as two consecutive PA time steps. Then the evolution  of the particle system in the time interval $[t_0,t_1)$ can be decomposed into two stages: (1) perform PA for the original weighted empirical measure $\nu_{t_0}$ and obtain a signed empirical measure $\widetilde{\nu}_{t_0}$; (2) evolve $\widetilde{\nu}_{t_0}$ following the law of the branching random walk and get another weighted empirical measure $\nu_{t_1}$. We denote the particles after PA endowed with position $\widetilde{\vb*{x}}_i(t)$, wave vector $\widetilde{\vb*{k}}_i(t)$ and weight $\widetilde{w}_i(t)$.

% error estimation between two annihilation strategies
\begin{proof}[Proof of Theorem~\ref{thm:main theorem}]
Following the flowchart in Fig.~\ref{fig:flowchart}, we take our first step by considering the error between two consecutive PA moments $t_1>t_0$. We hope to estimate $L^2$ error $\mathbb{E}[\|\mathcal{P}_h\nu_t-\mathcal{P}_h f(t)\|_2^2 \big| \nu_{t_0}]$ at time $t\in(t_0,t_1]$. By the projection identity Eq. \eqref{eq:projection identity} and the weak error definition
\begin{equation}
    \label{eq:weak error definition}
    e_t(\varphi_t) = \langle \varphi_t, \nu_{t} \rangle - \langle \varphi_t, f(t)\rangle, 
\end{equation}
we can express the $L^2$ error as a sum of weak errors for each piecewise constant basis function $\varphi_k$, that is,
\begin{equation}
    \label{eq: L2 error identity}
    \mathbb{E}[\|\mathcal{P}_h\nu_t-\mathcal{P}_h f(t)\|_2^2 \big| \nu_{t_0}] = \sum_k |Q_k|\mathbb{E}[e_t^2(\varphi_k)\big| \nu_{t_0}].
\end{equation}
Then it turns to estimate the weak error $\mathbb{E}[e_t^2(\varphi_t)\big| \nu_{t_0}]$ for suitable test function $\varphi_t$. By the bias-variance decomposition $\mathbb{E}[X^2] = (\mathbb{E}[X])^2+\Var(X)$ and law of total variance $\Var(X) = \Var(\mathbb{E}[X|Y])+\mathbb{E}[\Var(X|Y)]$, we can further decompose the weak error
\begin{equation}
    \label{eq:weak error decomposition}
    \mathbb{E}\left[e_t^2(\varphi_t)\big| \nu_{t_0}\right] =: I_1^t + I_2^t + I_3^t, 
\end{equation}
where the first term $I_1^t$ represents propagation error, the second term $I_2^t$ represents the variance from residual resampling, and the third term $I_3^t$ represents the variance from WBRW, that is, 
\begin{equation}
    \label{eq:weak error decomposition parts}
    \begin{aligned}
        I_1^t &= \left(\mathbb{E}\left[\langle \varphi_t, \nu_{t} \rangle\big| \nu_{t_0} \right]-\langle \varphi_t, f(t)\rangle\right)^2, \\
        I_2^t &= \Var\left(\mathbb{E}\left[\langle \varphi_t, \nu_{t} \rangle\big| \widetilde{\nu}_{t_0}, \nu_{t_0}\right]\big| \nu_{t_0}\right) =\Var\left(\mathbb{E}\left[\langle \varphi_t, \nu_{t} \rangle\big| \widetilde{\nu}_{t_0}\right]\big| \nu_{t_0}\right), \\
        I_3^t &= \mathbb{E}\left[\Var\left(\langle \varphi_t, \nu_{t} \rangle\big| \widetilde{\nu}_{t_0}, \nu_{t_0}\right)\big| \nu_{t_0}\right]=\mathbb{E}\left[\Var\left(\langle \varphi_t, \nu_{t} \rangle\big| \widetilde{\nu}_{t_0}\right)\big| \nu_{t_0}\right].
    \end{aligned}
\end{equation}
Here the second equalities in $I_2^t$ and $I_3^t$ are due to the Markovian property of the branching random walk. In the following, we will estimate each term in Eq. \eqref{eq:weak error decomposition parts} and then combine them to iterate and prove Theorem \ref{thm:main theorem}.

% I_1^t bias using L^2 stability
\paragraph{As for $I_1^t$} Let $U(t,s)$ be the evolution operator for the Wigner equation with time-dependent generator $\mathcal{L}(t)=-\frac{\hbar \vb*{k}}{m}\cdot\nabla_x+\Theta_{V(t)}$, so that $f(t)=U(t,s)f(s)$ and $\varphi(s)=U(t,s)^*\varphi_t$. Note that $\nu_t$ evolves from $\widetilde{\nu}_{t_0}$, which is sampled from $\mathcal{P}_h\nu_{t_0}$ by PAUM. Thus the unbiasedness in Eq.~\eqref{eq:unbiasedness} gives $\mathbb{E}\left[\langle \varphi_t, \nu_t \rangle\big| \nu_{t_0} \right]=\langle \varphi_t,U(t,t_0)\mathcal{P}_h\nu_{t_0}\rangle$. Combining unitarity with the projection identity Eq.~\eqref{eq:projection identity}, we obtain
\begin{equation}
    \label{eq:I_1}
    \begin{aligned}
        \sum_k |Q_k| I_1^t(\varphi_k) = & \sum_k |Q_k| \left(\langle \varphi_k, U(t,t_0)(\mathcal{P}_h\nu_{t_0}-f(t_0))\rangle\right)^2\\
        = & \|\mathcal{P}_h U(t,t_0)(\mathcal{P}_h\nu_{t_0}-f(t_0))\|_2^2 
        \le \|\mathcal{P}_h\nu_{t_0}-f(t_0)\|_2^2.
    \end{aligned}
\end{equation}

% I_2^t residual resampling variance
\paragraph{As for $I_2^t$} The unbiasedness in Eq.~\eqref{eq:unbiasedness} implies $I_2^t(\varphi_t)=\Var\left(\langle\varphi(t_0), \widetilde{\nu}_{t_0}\rangle\big|\nu_{t_0}\right)$, which represents the variance from residual resampling. Note that $\widetilde{\nu}_{t_0}$ is sampled from $\mathcal{P}_h\nu_{t_0}$. Denote $\mathcal{I}_k(t)$ and $\widetilde{\mathcal{I}}_k(t)$ as the set of indices of particles in $Q_k$ at time $t$ before and after PA, respectively, i.e., $\mathcal{I}_k(t)=\{i: (\vb*{x}_i(t),\vb*{k}_i(t))\in Q_k\}$ and $\widetilde{\mathcal{I}}_k(t)=\{i: (\widetilde{\vb*{x}}_i(t),\widetilde{\vb*{k}}_i(t))\in Q_k\}$. We then separate the randomness of positions and weights by the expresion of piecewise constant approximation, that is,
\begin{equation*}
    \begin{aligned}
        \langle \varphi(t_0), \widetilde{\nu}_{t_0}-\mathcal{P}_h\nu_{t_0} \rangle=&\sum_k \sum_{i\in\widetilde{\mathcal{I}}_k(t_0)} \frac{\widetilde{w}_i(t_0)}{N_\alpha}(\varphi(\widetilde{\vb*{x}}_i(t_0),\widetilde{\vb*{k}}_i(t_0),t_0)-\langle \varphi(t_0), \varphi_k\rangle)\\
        &+\sum_k \langle \varphi(t_0), \varphi_k\rangle \frac{1}{N_\alpha}(
            \sum_{i\in\widetilde{\mathcal{I}}_k(t_0)} \widetilde{w}_i(t_0) - \sum_{i\in\mathcal{I}_k(t_0)} w_i(t_0)).
    \end{aligned}
\end{equation*}
Since the locations and weights are conditionally independent given $\nu_{t_0}$, we can derive the following inequality
\begin{equation*}
    \begin{aligned}
        I_2^t(\varphi_t)=&\frac{1}{N_\alpha^2}\sum_k\mathbb{E}[
            \sum_{i\in\widetilde{\mathcal{I}}_k(t_0)} \widetilde{w}_i^2(t_0)\big|\nu_{t_0}]\Var_{\varphi_k}(\varphi(t_0))\\
        &+\frac{1}{N_\alpha^2}\sum_k (\langle \varphi(t_0), \varphi_k\rangle)^2\Var(
            \sum_{i\in\widetilde{\mathcal{I}}_k(t_0)} \widetilde{w}_i(t_0)\big|\nu_{t_0})\\
        \le & \frac{1}{N_\alpha}\langle |\mathcal{P}_h\nu_{t_0}|, (\varphi-\mathcal{P}_h\varphi)^2(t_0) \rangle + \frac{1}{4N_\alpha^2|Q|}\sum_k |Q_k|(\langle \varphi(t_0), \varphi_k\rangle)^2\\
        \le & \frac{\|\mathcal{P}_h\nu_{t_0}\|_\infty}{N_\alpha}\|(\mathcal{I}-\mathcal{P}_h)\varphi(t_0)\|_2^2 + \frac{\|\varphi_t\|_2^2}{4N_\alpha^2|Q|}
        \le \left(\frac{\|\mathcal{P}_h\nu_{t_0}\|_\infty}{N_\alpha}+\frac{1}{4N_\alpha^2|Q|}\right)\|\varphi_t\|_2^2,
    \end{aligned}
\end{equation*}
where $|\widetilde{w}_i(t_0)|=1$ turns the sum of squared weights into the sum of absolute weights, and the residual resampling procedure in Eq.~\eqref{eq:residual resampling} bounds the variance of the sum of weights in each bin by $1/4$. We then use $\langle f,g\rangle\le\|f\|_\infty\|g\|_1$ and the projection identity Eq.~\eqref{eq:projection identity}. The final step follows from $\|(\mathcal I-\mathcal P_h)\varphi(t_0)\|_2^2\le\|\varphi(t_0)\|_2^2=\|\varphi_t\|_2^2$. Therefore, taking $\varphi_t=\varphi_k$ and summing over $k$, we have
\begin{equation}
    \label{eq:I_2}
    \sum_k |Q_k| I_2^t(\varphi_k) \le \frac{\|\mathcal{P}_h\nu_{t_0}\|_\infty|\Omega|}{N_\alpha|Q|} + \frac{|\Omega|}{4N_\alpha^2|Q|^2}.
\end{equation}

% I_3^t WBRW variance which grows exponentially with time
\paragraph{As for $I_3^t$} The term $I_3^t$ represents the variance from WBRW. Recall the exponential variance bound for WBRW in Proposition~\ref{prop:without PA} and consider the conditional independence of particles in WBRW, and conditional independence of positions and weights in PAUM, as well as the resampling weights $|\widetilde{w}_i(t_0)|=1$ for all particles. Then we can derive the following inequality
\begin{equation*}
    \begin{aligned}
        I_3^t(\varphi_t)=&\mathbb{E}[\Var(\langle\varphi_t,\nu_t\rangle\big|\widetilde{\nu}_{t_0})\big|\nu_{t_0}]=\mathbb{E}[\sum_i \frac{\widetilde{w}_i^2(t_0)}{N_\alpha^2}\Delta \Phi_{w}^{(2)}(\widetilde{\vb*{x}}_i(t_0),\widetilde{\vb*{k}}_i(t_0),t_0)\big|\nu_{t_0}]\\
        =&\frac{1}{N_\alpha}\sum_k\mathbb{E}[\frac{\sum_{i\in\widetilde{\mathcal{I}}_k(t_0)}|\widetilde{w}_i(t_0)|}{N_\alpha}]\langle \Delta \Phi_{w,t}^{(2)}(t_0),\varphi_k \rangle\\
        =&\frac{1}{N_\alpha}\sum_k |Q_k||\langle \nu_{t_0}, \varphi_k \rangle| \langle \Delta \Phi_{w,t}^{(2)}(t_0),\varphi_k \rangle=\frac{1}{N_\alpha}\langle |\mathcal{P}_h\nu_{t_0}|, \Delta \Phi_{w,t}^{(2)}(t_0)\rangle\\
        \le& \frac{\|\mathcal{P}_h\nu_{t_0}\|_\infty}{N_\alpha}\|\Delta \Phi_{w,t}^{(2)}(t_0)\|_1 \le \frac{\|\mathcal{P}_h\nu_{t_0}\|_\infty}{N_\alpha}2(\me^{\int_{t_0}^t 2\frac{\breve{\xi}^2(s)}{\gamma_0(s)}\D s}-1)\|\varphi_t\|_2^2.
    \end{aligned}
\end{equation*}
Let $\varphi_t=\varphi_k$ and sum over $k$, we have
\begin{equation}
    \label{eq:I_3}
    \sum_k |Q_k| I_3^t(\varphi_k) \le 2(\me^{\int_{t_0}^t 2\frac{\breve{\xi}^2(s)}{\gamma_0(s)}\D s}-1)\frac{\|\mathcal{P}_h\nu_{t_0}\|_\infty|\Omega|}{N_\alpha|Q|}.
\end{equation}
Combining \eqref{eq: L2 error identity} and \eqref{eq:weak error decomposition} with \eqref{eq:I_1}-\eqref{eq:I_3} gives the one-step error bound, we can obtain the error estimation between two consecutive PA time steps,
\begin{equation}
    \begin{aligned}
        \mathbb{E}\left[
        \|\mathcal{P}_h\nu_t-f(t)\|_2^2\big|\nu_{t_0}
        \right] \le & \|\mathcal{P}_h \nu_{t_0}-f(t_0)\|_2^2+\frac{\|\mathcal{P}_h\nu_{t_0}\|_\infty|\Omega|}{N_\alpha|Q|}+\frac{|\Omega|}{4N_\alpha^2|Q|^2}\\
        &+\|(\mathcal{I}-\mathcal{P}_h)f(t)\|_2^2+ 2(\me^{\int_{t_0}^t 2\frac{\breve{\xi}^2(s)}{\gamma_0(s)}\D s}-1)\frac{\|\mathcal{P}_h \nu_{t_0}\|_\infty|\Omega|}{N_\alpha|Q|}.
        \label{eq:wp-error-step}
    \end{aligned}
\end{equation}
Here the projection error identity $\|\varphi\|_2^2=\|\mathcal{P}_h\varphi\|_2^2+\|(\mathcal{I}-\mathcal{P}_h)\varphi\|_2^2$ is used with $\varphi=\mathcal{P}_h\nu_t-f(t)$. The same orthogonality gives the initial error decomposition
\begin{equation*}
    \|\mathcal P_h\nu_0-f(0)\|_2^2
    =\|\mathcal P_h\nu_0-\mathcal P_h f(0)\|_2^2
    +\|(\mathcal I-\mathcal P_h)f(0)\|_2^2.
\end{equation*}
Taking expectations conditional on the PA schedule and iterating Eq.~\eqref{eq:wp-error-step} up to $T$ yields Theorem~\ref{thm:main theorem}, with the initial sampling term included in $\mathcal E_r$ and the initial projection term included in $\mathcal E_p$.
\end{proof}

\subsection{Analytical solution of the optimization problem~\eqref{eq:choice mixed cost}}
\begin{proof}[Proof of Proposition~\ref{prop:choice optimum}]
We first minimize \eqref{eq:choice mixed cost} over $y$ at fixed $\rho$, then minimize the resulting cost over $\rho\geq1$. At fixed $\rho$, the sign of $\partial_y\mathcal C$ is the sign of
\begin{equation}
\label{eq:yC}
 D(y,\rho)=\eta e^y(y-1)+(1-\eta)e^{\rho y}(\rho y-1),
\end{equation}
which is strictly increasing in $y>0$ due to its derivative $\eta ye^y+(1-\eta)\rho^2ye^{\rho y}>0$. Hence by bisection, there is a unique root $y_C(\rho)\in[\frac{1}{\rho},1]$ of \eqref{eq:yC}. The cost decreases before $y_C(\rho)$ and increases afterward, so the best admissible PA time step given $\rho$ is
\begin{equation}
\label{eq:choice fixed rho optimum}
 y^*(\rho)=\min\{\rho x_V,y_C(\rho)\}.
\end{equation}

Here we discuss the cases where $y^*=\rho x_V$ and $y^*=y_C(\rho)$ in sequence, and then combine them to find the joint optimum. First, consider the branch $y_C(\rho)\geq \rho x_V$ and $y_*(\rho)=\rho x_V$. The computational cost per unit $y$ becomes
\begin{equation}
\label{eq:choice variance branch}
 J_V(\rho):=\mathcal C(\rho x_V,\rho)
 =\frac{\eta e^{\rho x_V}+(1-\eta)e^{\rho^2x_V}}{\rho x_V}.
\end{equation}
Its derivative has the sign of
\begin{equation}
\label{eq:choice phi}
 \Phi(\rho)=\eta e^{\rho x_V}(\rho x_V-1)
 +(1-\eta)e^{\rho^2x_V}(2\rho^2x_V-1).
\end{equation}
Moreover, $\Phi'(\rho)=\eta\rho x_V^2e^{\rho x_V}+2(1-\eta)\rho x_Ve^{\rho^2x_V}(2\rho^2x_V+1)>0$. Thus $\Phi$ is strictly increasing and has at most one zero $\rho_\Phi$. Note that $\Phi(1)=e^{x_V}[(2-\eta)x_V-1]$. When $x_V<(2-\eta)^{-1}$ and $\Phi(1)<0$, this branch has a unique stationary point $\rho_\Phi>1$ and the cost $J_V(\rho)$ strictly decreases before $\rho_\Phi$ and increases afterward. When $x_V\geq(2-\eta)^{-1}$, $\Phi(1)\geq0$ and the cost $J_V(\rho)$ strictly increases for $\rho\geq1$.

On the other branch $y_C(\rho)\le \rho x_V$ and $y_*(\rho)=y_C(\rho)$, we denote $J_C(\rho):=\mathcal C(y_C(\rho),\rho)$. The derivative of $J_C$ with respect to $\rho$ is simplified by the chain rule and the fact that $\partial_y\mathcal C(y_C(\rho),\rho)=0$, that is, 
\begin{equation*}
    \frac{\dd}{\dd \rho}J_C(\rho) = \partial_y \mathcal C(y_C(\rho),\rho) \frac{\dd y_C}{\dd \rho} + \partial_\rho \mathcal C(y_C(\rho),\rho) = \partial_\rho \mathcal C(y_C(\rho),\rho)=(1-\eta)e^{\rho y_C(\rho)}\geq0.
\end{equation*}
Thus the minimum in the branch occurs at the left endpoint $\rho=\max(1, \rho_c)$, where the crossing point $\rho_c$ is the minimum solution of $y_C(\rho_c)=\rho_c x_V$.

In summary, the optimal $\rho$ is chosen from $\rho_c$, $\rho_\Phi$ and $1$. We claim that the cost at $\rho_c$ is never smaller than the cost at $\rho_\Phi$ or $1$. At a crossover $\rho_c>1$, its cost can also be expressed as $J_V(\rho_c)$, since $y_C(\rho_c)=\rho_c x_V$. The condition $D(y_C(\rho_c),\rho_c)=0$ in Eq.~\eqref{eq:yC} and the monotonicity of $\Phi$ implies
\begin{equation}
\label{eq:choice crossing condition}
 \Phi(\rho_c)=(1-\eta)e^{\rho_c^2x_V}\rho_c^2x_V\geq 0=\Phi(\rho_{\Phi}) \quad\implies \rho_c\geq \rho_{\Phi},\quad J_V(\rho_c)\geq J_V(\rho_{\Phi}).
\end{equation}
Thus when $\rho_\Phi>1$, the optimal branching ratio is $\rho^*=\rho_\Phi$, otherwise $J_V(\rho)$ is strictly increasing and the optimal ratio is $\rho^*=1$, which implies $y_C(\rho^*)=1$. These observations give the joint optimum.
\end{proof}

\section{Numerical experiments}
\label{sec:as}
We present numerical experiments to validate the theoretical effect of PA on the accuracy and long-time behavior, and to demonstrate the efficiency of the proposed scheduling. We compare the numerical Wigner function with the advective-spectral-mixed method reference solution~\cite{Xiong2016SEM} using the relative error $\mathcal E_2[f]=\|f^{\mathrm{num}}-f^{\mathrm{ref}}\|_2/\|f^{\mathrm{ref}}\|_2$ with $N_x=400, N_k=200$. For the two-body problem, $f$ is the same reduced Wigner function as in~\cite{Shao2019,Xiong2016SEM}.

All the numerical results are obtained with our own C++ serial implementations on a desktop computer equipped with an AMD Ryzen Threadripper 2990WX processor (3.0 GHz; 32 CPU cores) and 96 GB memory.

\subsection{Gaussian barrier scattering}
\label{sec:as1}
First, we consider typical 2D Gaussian barrier scattering problem~\cite{Shao2019} as a benchmark. The barrier and its Wigner kernel read
\begin{equation*}
  V(x)=H_B e^{-(x-x_B)^2/2},\quad
  V_W(x,k)=\frac{2H_B}{\hbar}\sqrt{\frac{2}{\pi}}\,e^{-2k^2}\sin\bigl(2k(x-x_B)\bigr),
\end{equation*}
where $H_B=0.3\,\mathrm{eV}$, $x_B=0\,\mathrm{nm}$ and more details refer to~\cite{Shao2019}. Here Theorem~\ref{thm:main theorem} reduces to the time-independent case, where $\breve{\xi}(t)\equiv\breve{\xi}$ and $\gamma_0(t)\equiv\gamma_0$ are constants. Hence the variance bound $\mathcal E_w$ is simplified to $\sum_{r=0}^{R-1} 2(\me^{2\breve{\xi}^2(t_{r+1}-t_r)/\gamma_0}-1)\mathbb{E}[\|\mathcal{P}_h \nu_{t_r}\|_\infty|\Omega|]/(N_\alpha|Q|)$, and the mapping between the intrinsic time and physical time is $\theta(t)=2\breve{\xi}^2 t/\gamma_0$. For ten independent runs $f^{(r)}$ with mean $\bar f$, we estimate the numerical variance by $\widehat{\mathrm{Var}}=\frac{1}{9}\sum_{r=1}^{10}\|f^{(r)}-\bar f\|_2^2$ and the numerical squared bias by $\widehat{\mathrm{Bias}^2}=\|\bar f-f^{\mathrm{ref}}\|_2^2-\widehat{\mathrm{Var}}/10$. 

The upper panel of Figure~\ref{fig:benchmark-comparison} illustrates that ratio of the numerical variance to the estimated one of Theorem~\ref{thm:main theorem}, $\widehat{\mathrm{Var}}/\mathcal E_w$, stays between $0.01$ and $0.001$ over time and becomes stationary after a certain time. Thus $\mathcal E_w$ tracks variance growth up to a scale factor for PAUM, and also for SPADE and SPADE-Mix, verifying the analogy in Remark~\ref{rmk:analogy}. The insets show the linear growth of numerical variance, confirming the effect of PA in suppressing variance growth. The lower panel of Figure~\ref{fig:benchmark-comparison} shows that as $T_A$ grows, squared bias decreases while variance increases, and MSE attains a minimum at an intermediate $T_A$. Figure~\ref{fig:adaptive-efficiency} shows that our scheduling strategy allows all three PA methods to populate the low-cost regime of the empirical Pareto frontier, approaching fine-tuned uniform cost-error trade-offs without prior knowledge of $T_A$. 

\begingroup
\setlength{\intextsep}{3pt plus 1pt minus 1pt}
\setlength{\abovecaptionskip}{3pt}
\captionsetup[subfloat]{farskip=2pt,captionskip=2pt}
\begin{figure}[H]
  \centering
  \subfloat[PAUM\label{fig:paum-variance}]{%
    \includegraphics[width=0.315\textwidth]{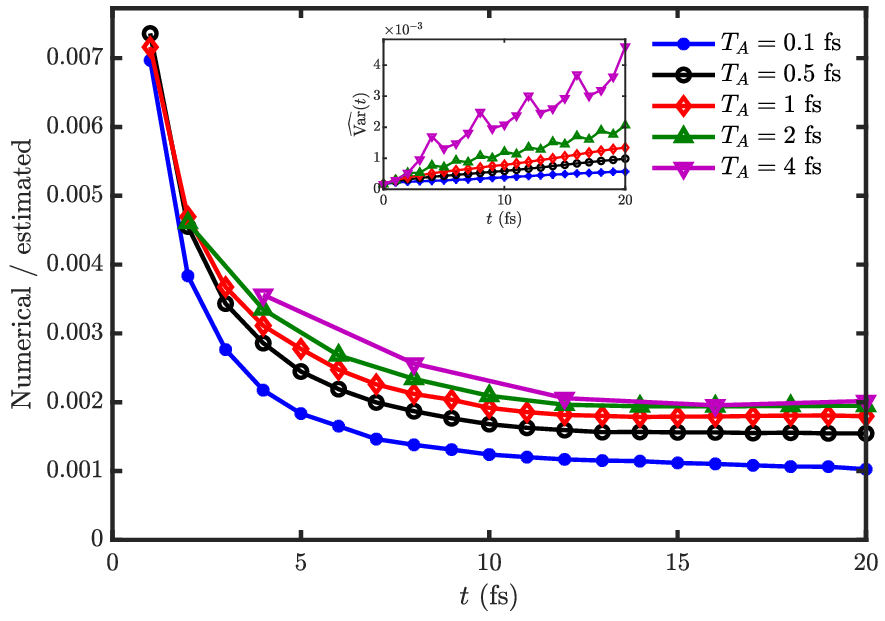}}\hfill
  \subfloat[SPADE\label{fig:spade-variance}]{%
    \includegraphics[width=0.315\textwidth]{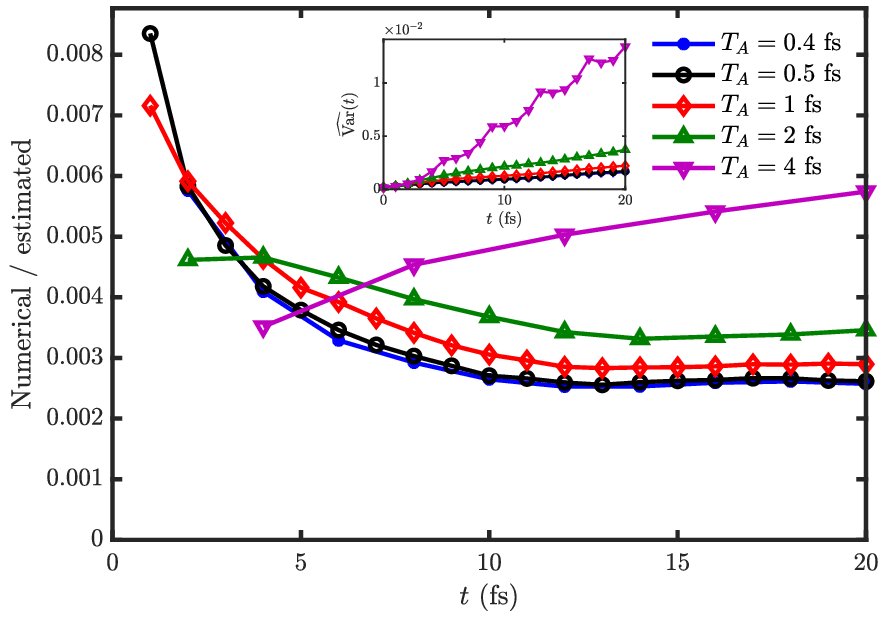}}\hfill
  \subfloat[SPADE-Mix\label{fig:spade-mix-variance}]{%
    \includegraphics[width=0.315\textwidth]{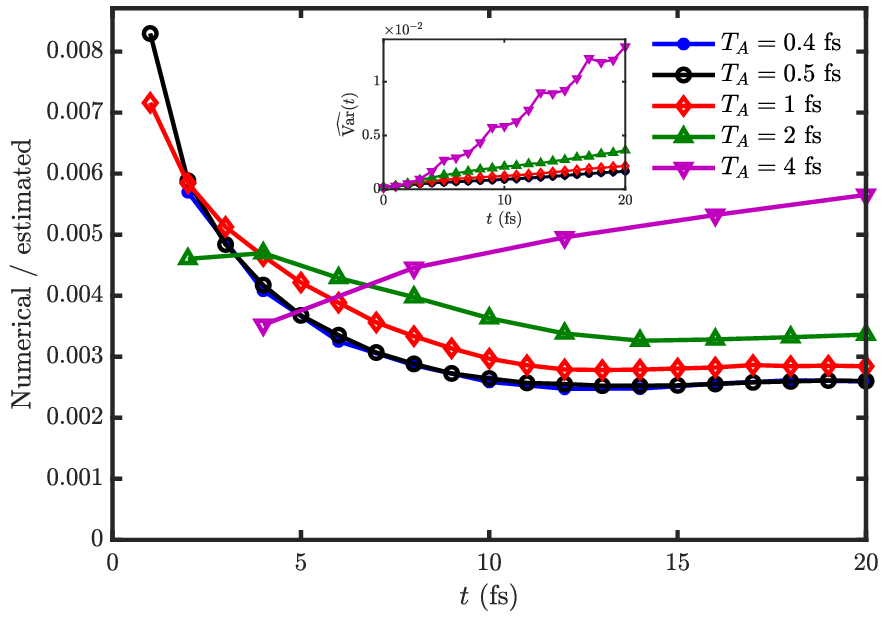}}
  \par
  \subfloat[PAUM\label{fig:paum-competition}]{%
    \includegraphics[width=0.315\textwidth]{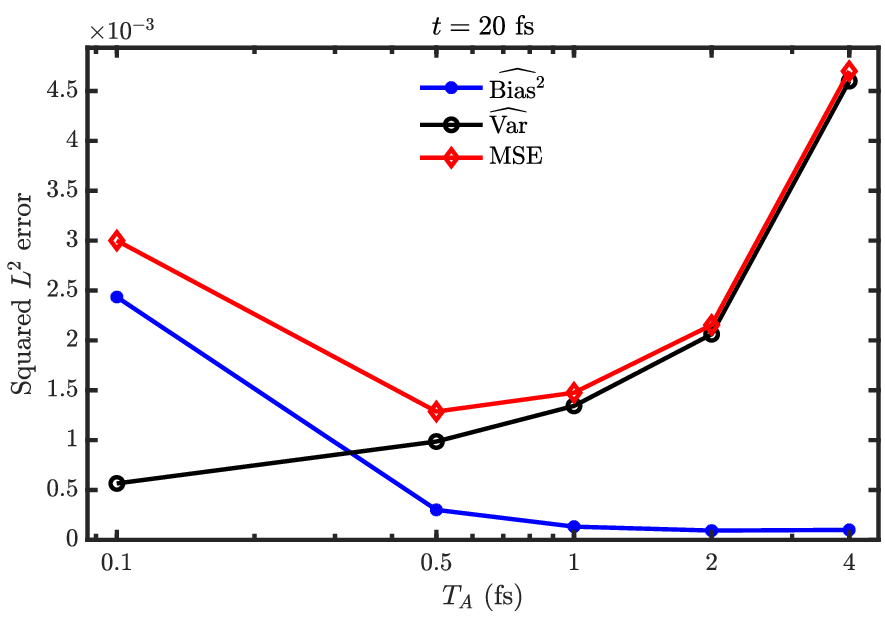}}\hfill
  \subfloat[SPADE\label{fig:spade-competition}]{%
    \includegraphics[width=0.315\textwidth]{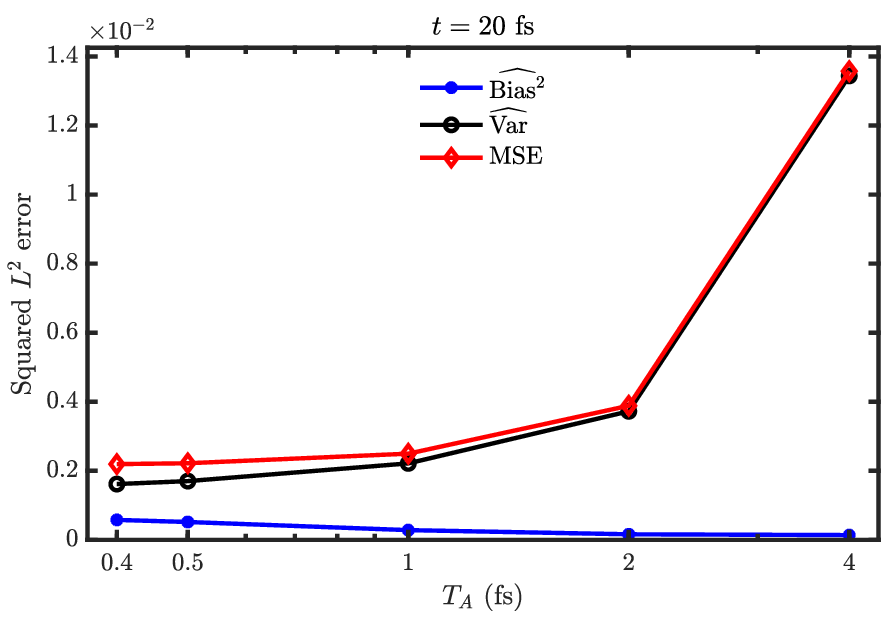}}\hfill
  \subfloat[SPADE-Mix\label{fig:spade-mix-competition}]{%
    \includegraphics[width=0.315\textwidth]{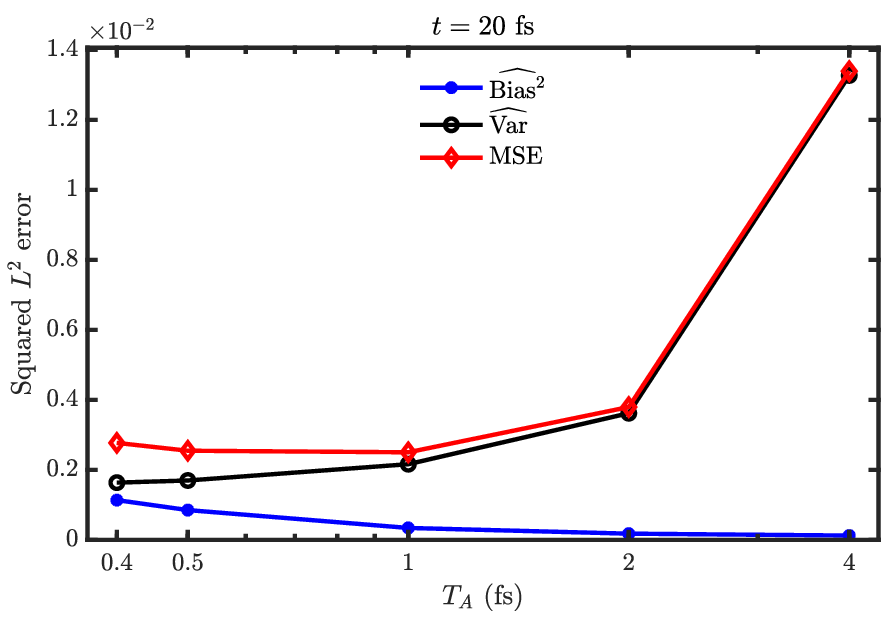}}
  \caption{2D Gaussian barrier scattering at $T=20$ fs and $N_\alpha=1\times 10^6$ with different fixed $T_A$: (top) Squared bias, variance and their sum at final time $T$; (bottom) The ratio of numerical variance to estimated one over time, and insets show the linear growth of numerical variance. }
  \label{fig:benchmark-comparison}
\end{figure}

\begin{figure}[H]
  \centering
  \subfloat[PAUM\label{fig:paum-efficiency}]{%
    \includegraphics[width=0.315\textwidth]{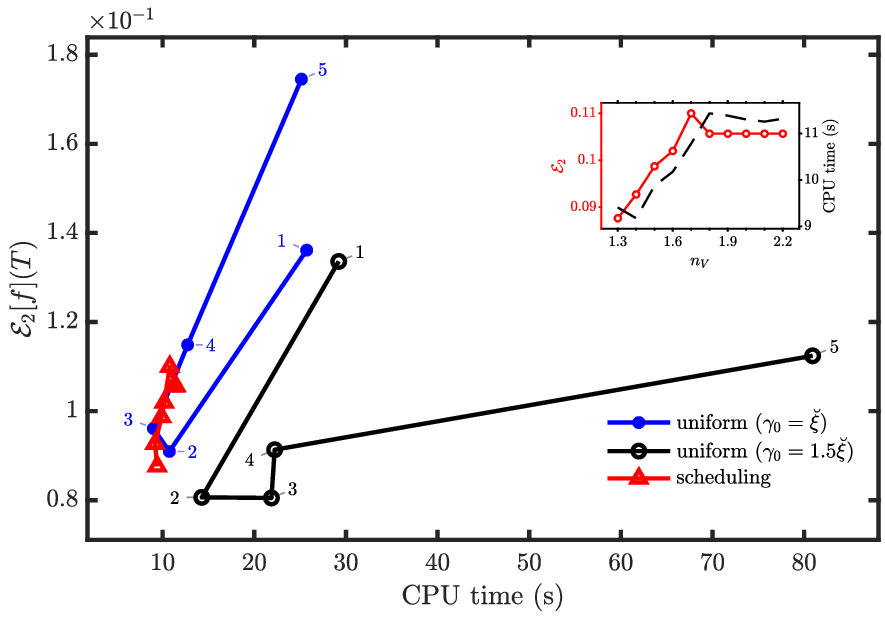}}\hfill
  \subfloat[SPADE\label{fig:spade-efficiency}]{%
    \includegraphics[width=0.315\textwidth]{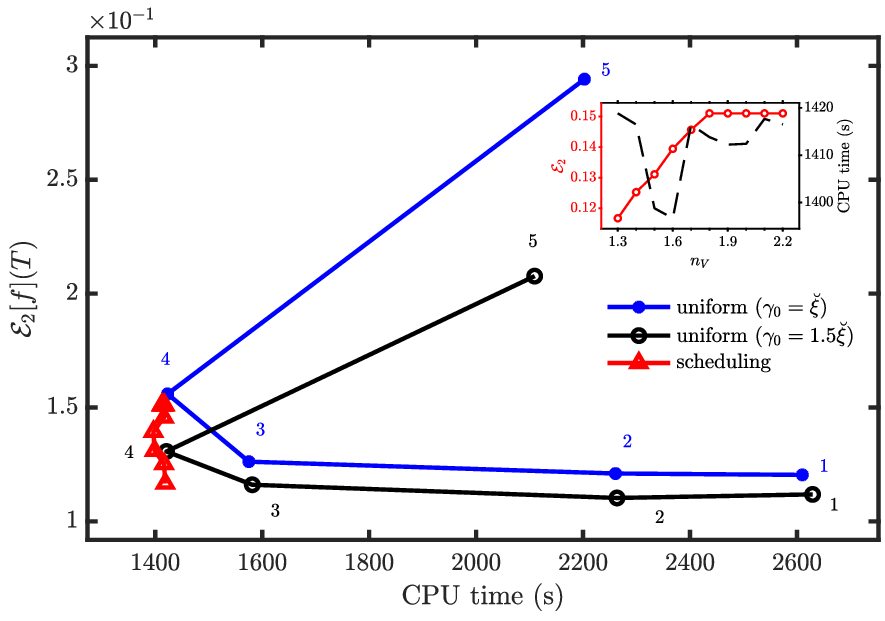}}\hfill
  \subfloat[SPADE-Mix\label{fig:spade-mix-efficiency}]{%
    \includegraphics[width=0.315\textwidth]{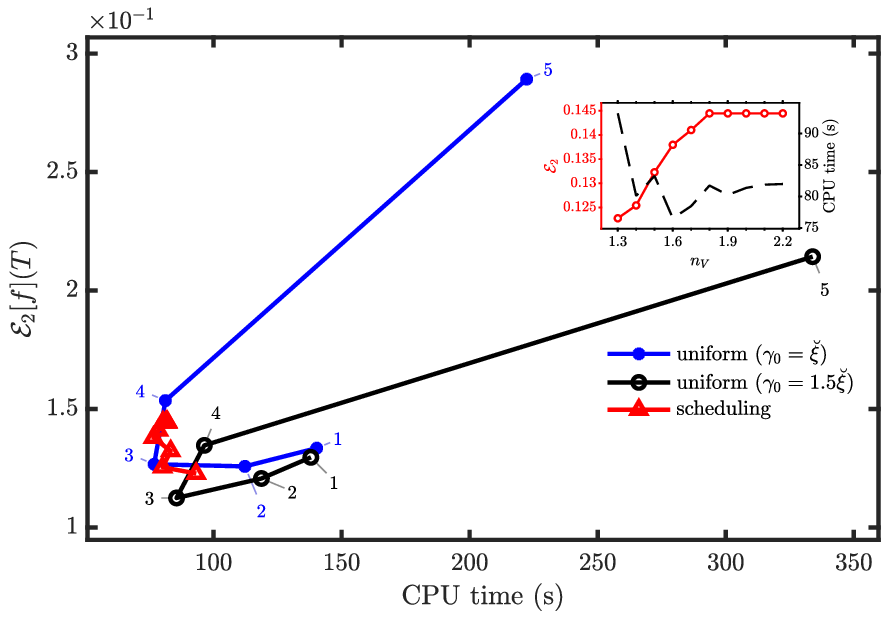}}
  \caption{2D Gaussian barrier scattering at $T=20$ fs and $N_\alpha=1\times 10^6$: $\mathcal{E}_2[f](T)$ versus CPU time. For each PA method, the performance of scheduling reach the Pareto frontier of uniform strategies. Label $1$-$5$ denote $T_A=0.1,0.5,1.0,2.0,4.0$ fs for PAUM, and $T_A=0.4,0.5,1.0,2.0,4.0$ fs for SPADE and SPADE-Mix, respectively.}
  \label{fig:adaptive-efficiency}
\end{figure}

\subsection{Time-dependent double barrier}
We consider a time-dependent double barrier scattering problem~\cite{Xiong2019} under potential
\begin{equation*}
  V(x,t) = 0.8\bigl(0.5-0.5\cos(0.1t)\bigr)e^{-(x+10)^2/8}+0.8e^{-(x-10)^2/8}.
\end{equation*}

Relative to $T_A=1$ fs interval used in~\cite{Xiong2019}, the proposed scheduling strategy speeds up the simulation $3.38\times$ with a $1.9\%$ increase in final relative error as shown in Table~\ref{tab:tdg-efficiency}. In Figure~\ref{fig:tdg-adaptive-results}(b), PA intervals $T_A(t)$, which is adaptively computed by Algorithm~\ref{optimal alg}, fluctuates around its average $0.42$ fs adaptively, lengthening when $\breve{\xi}(t)$ is small and shortening when $\breve{\xi}(t)$ is large. 

Using the average value of $T_A$ from the adaptive schedule, we also test a uniform $T_A=0.42$ fs schedule, which gives a $3.77\times$ speedup with comparable final error. That is, the adaptively selected intervals indeed provide a useful reference for choosing an efficient uniform schedule. Although this fixed schedule is faster in the present example, Figure~\ref{fig:tdg-adaptive-results}(c) demonstrates that our scheduling strategy yields the lowest peak particle count before PA, and the peak is approximately one tenth of that of the $1$ fs baseline, significantly reducing the maximum storage required for particles.

Beyond decreasing memory usage, adaptive scheduling responds to temporal changes in branching activity that a fixed average interval cannot capture. To illustrate why this distinction matters, consider a hypothetical experiment in which the potential vanishes over an long interval and then branching is switched off. Adaptive scheduling can then skip PA since it is calculated under intrinsic time $\theta(t)$ in Eq.~\eqref{eq:intrinsic theta}, whereas a fixed-frequency schedule continues to incur unnecessary annihilation costs, illustrating the efficiency advantage of adapting to the time-dependent interaction strength.

\begin{table}[H]
  \centering
  \caption{2D time-dependent barrier under PAUM at $T=60$ fs and $N_\alpha=4\times 10^6$: efficiency comparison. Scheduling gives a $3.38\times$ speedup over the $1$ fs baseline. By averaging the adaptive intervals, our algorithm recommends a fixed PA interval of $T_A=0.42$ fs, which gives a $3.77\times$ speedup. Both schedules yield comparable final relative errors.} 
  \label{tab:tdg-efficiency}
\begin{tabular}{lrrr}
\hline
Parameters & CPU time (s) & $\mathcal{E}_2[f](T)$ & Speedup \\
\hline
$T_A=1$ fs, $\gamma_0=\max_{t\in[0,T]}\breve{\xi}(t)$ & 3259.61 & \textbf{0.29778} & 1.00$\times$ \\
scheduling ($n_V=2$) & 963.15 & 0.30345 & 3.38$\times$ \\
$T_A=0.42$ fs, $\gamma_0=\max_{t\in[0,T]}\breve{\xi}(t)$ & \textbf{863.72} & 0.30076 & \textbf{3.77}$\times$ \\
\hline
\end{tabular}
\end{table}

\begin{figure}[H]
  \centering
  \subfloat[Relative $L^2$ error\label{fig:tdg-error}]{%
    \includegraphics[width=0.315\textwidth]{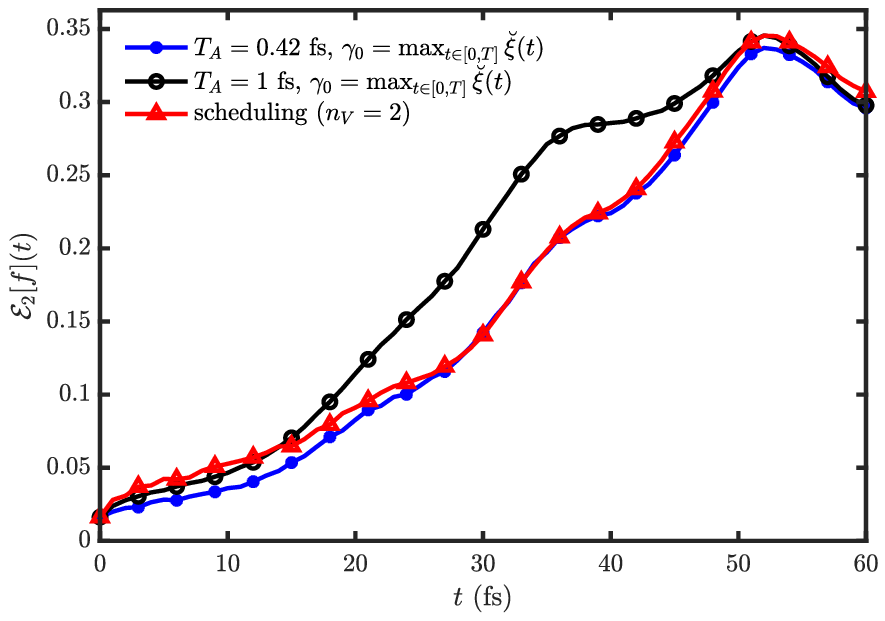}}\hfill
  \subfloat[Annihilation interval $T_A$\label{fig:tdg-interval}]{%
    \includegraphics[width=0.315\textwidth]{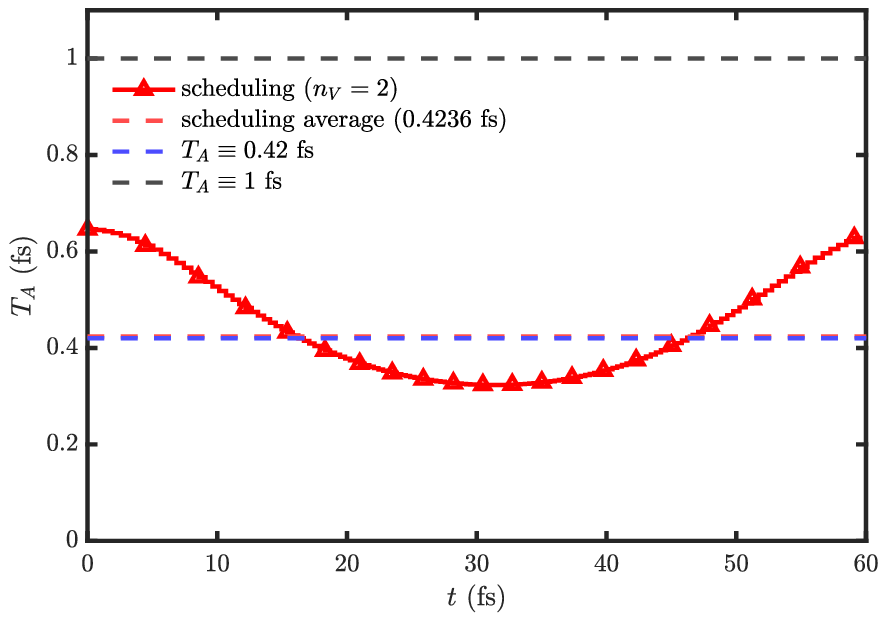}}\hfill
  \subfloat[Particle counts\label{fig:tdg-particles}]{%
    \includegraphics[width=0.315\textwidth]{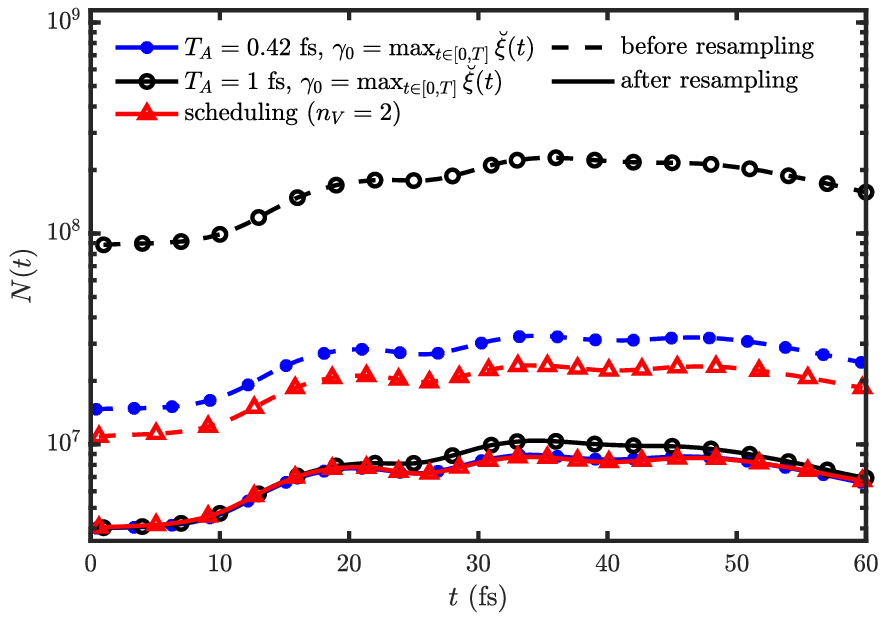}}
  \caption{2D time-dependent barrier under PAUM at $T=60$ fs and $N_\alpha=4\times 10^6$. (left) The relative errors remain comparable throughout the run. (middle) Scheduling shortens $T_A$ as $\breve{\xi}(t)$ rises. (right) Our method also gives the lowest peak particle count before resampling.}
  \label{fig:tdg-adaptive-results}
\end{figure}

\section{Conclusion and outlook}
\label{sec:con}
The theoretical analysis in this work provides a quantitative foundation for adaptive PA scheduling in long-time WBRW simulations. Our analysis uses the Markov property of the full particle system and the conditional independence of the branching evolutions initiated by distinct particles between successive PA steps. These properties allow us to track how branching variance, resampling variance, and annihilation bias contribute to error accumulation. Building on this error structure, a per-unit-time cost formulation yields a variance-constrained optimization problem for selecting PA intervals and branching rates. An intrinsic-time reparameterization provides a common clock for variance control, PA scheduling, and branching dynamics under time-dependent potentials. Together, these ingredients connect the theoretical characterization of long-time error accumulation with efficient scheduling strategy in practice. 

There remain several aspects to be explored. The most important one is to extend the current framework to realistic problems in higher dimensions, in combined with variance reduction techniques, such as stationary phase approximation in~\cite{Shao2020} and  increasing the lower bound of $\gamma_0/\breve{\xi}$. With the speedup of the proposed scheduling strategy, our ongoing work promisingly targets the simulation in the 6-, 12-, and 18-dimensional phase spaces. Also, developing computable bias estimates for SPADE and SPADE-Mix is a further direction, which helps incorporate direct control of accumulated annihilation bias into the scheduling criteria.

\section*{Acknowledgement}
This research was supported by the National Natural Science Foundation of China (Nos.~12325112, 12288101) and the High-performance Computing Platform of Peking University.

\end{document}